\documentclass[useAMS,usenatbib]{mn2e}

\usepackage{graphicx}
\usepackage{amsmath}
\usepackage{amssymb}
\usepackage{bm}
\usepackage{ulem}

\newcommand{\be}{\begin{equation}}
\newcommand{\ee}{\end{equation}}
\newcommand{\bsplit}{\begin{split}}
\newcommand{\Exp}[1]{{\rm e}^{#1}}
\newcommand{\del}{\partial}
\newcommand{\Del}{{\nabla}}
\newcommand{\bmDel}{\bm{\nabla}}

\newcommand{\bfu}{\bm{u}}
\newcommand{\bmb}{\bm{b}}

\newcommand{\alp}{\alpha}

\newcommand{\Emf}{\bm{\mathcal{E}}}
\newcommand{\Flux}{\bm{\mathcal{F}}}
\newcommand{\Fmf}{\bm{F}}
\newcommand{\Emfr}{\mathcal{E}_r}
\newcommand{\Emfphi}{\mathcal{E}_\phi}
\newcommand{\Emfz}{\mathcal{E}_z}

\newcommand{\Fmfr}{F_r}
\newcommand{\Fmfphi}{F_\phi}
\newcommand{\Fmfz}{F_z}
\newcommand{\mean}[1]{\overline{#1}}
\newcommand{\meanv}[1]{\overline{\bm{#1}}}
\newcommand{\corr}{_\mathrm{c}}					
\newcommand{\corot}{_\mathrm{cor}}				
\newcommand{\D}{_\mathrm{D}}					
\newcommand{\eq}{_\mathrm{eq}}					
\newcommand{\rms}{_\mathrm{0}}					
\newcommand{\f}{_\mathrm{0}}				   	
\newcommand{\kin}{_\mathrm{k}}			   		
\newcommand{\magn}{_\mathrm{m}}			   		
\newcommand{\turb}{_\mathrm{t}}			   		
\newcommand{\crit}{\mathrm{cr}}			   		
\newcommand{\const}{\mathrm{const}}		   		
\newcommand{\ma}{_\mathrm{max}}			   		

\newcommand{\on}{_\mathrm{on}}
\newcommand{\off}{_\mathrm{off}}
\newcommand{\cro}{\times}
\newcommand{\Rm}{\mathcal{R}_\mathrm{m}}
\newcommand{\mbr}{\mean{B}_r}
\newcommand{\mbp}{\mean{B}_\phi}
\newcommand{\mbz}{\mean{B}_z}
\newcommand{\mbi}{\mean{B}_i}
\newcommand{\mur}{\mean{U}_r}
\newcommand{\mup}{\mean{U}_\phi}
\newcommand{\muz}{\mean{U}_z}

\newcommand\bgreek[1]{ \mathchoice
    {\hbox{\boldmath$\displaystyle{#1}$\unboldmath}}%
    {\hbox{\boldmath$\textstyle{#1}$\unboldmath}}%
    {\hbox{\boldmath$\scriptstyle{#1}$\unboldmath}}%
    {\hbox{\boldmath$\scriptscriptstyle{#1}$\unboldmath}}}
    
  \newcommand{\kms}{\,{\rm km\,s^{-1}}}
  \newcommand{\cmcms}{\,{\rm cm^2\,s^{-1}}}
  \newcommand{\kmskpc}{\,{\rm km\,s^{-1}\,kpc^{-1}}}

  \newcommand{\kpc}{\,{\rm kpc}}

  \newcommand{\Myr}{\,{\rm Myr}}
  \newcommand{\Gyr}{\,{\rm Gyr}}

\title[Galactic spiral patterns and dynamo action I]{Galactic spiral patterns and dynamo action I:\\ A new twist on magnetic arms}
\author[L.~Chamandy, K.~Subramanian and A.~Shukurov]{Luke Chamandy$^{1}$, Kandaswamy Subramanian$^{1}$ \& Anvar Shukurov$^{2,1}$
\thanks{E-mail: lchamandy@gmail.com (LC); kandu@iucaa.ernet.in (KS); anvar.shukurov@newcastle.ac.uk (AS)}\\
$^{1}$Inter-University Centre for Astronomy and Astrophysics, Post Bag 4, Ganeshkhind, Pune 411007, India\\
$^{2}$School of Mathematics \& Statistics, Newcastle University, Newcastle upon Tyne NE1 7RU}

\begin{document}

\date{Submitted to MNRAS}

\pagerange{\pageref{firstpage}--\pageref{lastpage}} \pubyear{2012}

\maketitle

\label{firstpage}


\begin{abstract}
We generalise the theory of mean-field galactic dynamos by allowing for temporal non-locality in the mean electromotive force (emf). 
This arises in random flows due to a finite response time of the mean emf to changes in the mean magnetic field and small-scale turbulence,
and leads to the telegraph equation for the mean field.
The resulting dynamo model also includes the nonlinear dynamo effects arising from magnetic helicity balance.
Within this framework, coherent large-scale magnetic spiral arms superimposed on the dominant axially symmetric magnetic structure are considered. 
A non-axisymmetric forcing of the mean-field dynamo by a spiral pattern (either stationary or transient) is invoked, 
with the aim of explaining the phenomenon of magnetic arms.
For a stationary dynamo forcing by a rigidly rotating material spiral, 
we find corotating non-axisymmetric magnetic modes enslaved to the axisymmetric modes and strongly peaked around the corotation radius.
For a forcing by transient material arms wound up by the galactic differential rotation, 
the magnetic spiral is able to adjust to the winding so that it resembles the material spiral at all times.
There are profound effects associated with the temporal non-locality, i.e. finite `dynamo relaxation time'.
For the case of a rigidly rotating spiral, 
a finite relaxation time causes each magnetic arm to mostly lag the corresponding material arm with respect to the rotation.
For a transient material spiral that winds up,
the finite dynamo relaxation time leads to a large, negative (in the sense of the rotation) 
phase shift between the magnetic and material arms, similar to that observed in NGC~6946 and other galaxies.
We confirm that sufficiently strong random seed fields can lead to global reversals of the regular field along the radius 
whose long-term survival depends on specific features of a given galaxy.
\end{abstract}


\section{Introduction}
\label{sec:intro}
Nearby disc galaxies are known to typically have regular (or large-scale or mean) magnetic fields
of 1--10\,$\mu$G in strength, which are coherent on the scale of the galaxies themselves \citep[see][ for reviews]{f10,be12}.
In many cases, galactic magnetic fields exhibit significant deviations from axial symmetry, 
in particular where the regular field is enhanced in `spiral magnetic arms' akin to the material arms. 
(By material arms, we specifically mean the regions where the densities of stars and gas are enhanced.)
In most cases, there is clearly a relationship between the material spiral arms and regular magnetic field spiral arms.
For instance, there exists a correspondence between the azimuthally averaged pitch angle of the regular field and that of the material spiral \citep{f10}.
This would be of little surprise, 
as the material arms naturally leave their imprint on the magnetic field, if not 
for the intriguing relation between magnetic and material arms, first discovered in the nearby galaxy
NGC~6946 \citep{behoernes96}. Here the magnetic arms are located almost precisely between the material arms. 
They appear to be phase-shifted images of the material arms, with a negative phase shift in the sense of the galactic rotation \citep{fricketal00, be07}. Importantly, the stronger tangling of the regular field by a
(presumably) more intense turbulence within the material arms cannot explain this phenomenon: both the
total and regular magnetic fields are stronger within the magnetic arms \citep{be07}. The nature
of magnetic arms remains to be convincingly explained.

\subsection{Magnetic arms and dynamo theory}
Turbulent mean-field dynamo theory has been successful in explaining the properties of the 
axisymmetric mode of regular fields in galaxies \defcitealias{rss88}{RSS88}\citep[][hereafter 
\citetalias{rss88}]{rss88}.
Although growing non-axisymmetric modes can arise in this theory (beyond a certain radius in the 
disc where the rotational velocity
shear is sufficiently small), they always have a lower growth rate than the axisymmetric mode
if the galactic disc is axially symmetric \citep{brss87,krss89}.
Thus, to explain non-axisymmetric modes as arising in an axisymmetric disc one must typically 
appeal to strongly non-axisymmetric seed fields and argue that the axisymmetric mode would not 
have had time to attain dominance.

An alternative, of course, is to appeal to the deviations of the galactic discs from axial 
symmetry. \defcitealias{ms91}{MS91} \citet[][hereafter \citetalias{ms91}]{ms91} and 
\defcitealias{sm93}{SM93} \citet[][hereafter \citetalias{sm93}]{sm93} explored analytically and 
numerically the growth of non-axisymmetric modes under an enhancement of dynamo action along a 
spiral (presumably, but not necessarily, co-spatial with the material spiral) with a constant global 
pattern speed. 
Subsequently,  \defcitealias{mo96}{M96}\citet[][hereafter \citetalias{mo96}; see also \citealt{mo98, moetal01}]{mo96} 
carried out mean-field dynamo simulations to explore the 
effects of modulating various quantities, such as the $\alpha$ effect, 
turbulent diffusivity, or the components of the mean velocity, along a spiral arm or bar.
Some authors have also addressed directly the question of how the regular field could become 
enhanced in between the material arms \citep{sh05, rohdeetal99}, though none of these explained the 
phenomenon of a substantial negative phase shift 
across a wide range of galactocentric distances, including those far away from
the corotation radius, as reported by \citet{fricketal00}.

\subsection{Advances in dynamo and spiral structure theories}\label{ADSST}
Much of the recent work on mean-field dynamo theory has been focussed on the nonlinear regime, 
and the possible catastrophic quenching of the dynamo implied by magnetic helicity conservation.
This has led to the development of the dynamical quenching theory 
\defcitealias{bs05a}{BS05}
\citep[see the review of][hereafter \citetalias{bs05a}]{bs05a}. 
In this theory, catastrophic quenching of the mean-field dynamo action is 
averted by a magnetic helicity flux, which transports small-scale magnetic helicity away from the region of dynamo action.
Dynamical quenching theory has been applied to local galactic dynamo models \citep{kleeorinetal00, vishniaccho01, kleeorinetal02, sssb06, sss07},  and to axisymmetric discs \citep{kleeorinetal02, smith12,smithetal12}, 
but not yet to non-axisymmetric mean-field disc dynamos. 
Past work on the non-axisymmetric disc dynamos focused on the linear (kinematic) regime, or relied on an approximate algebraic quenching formalism.

Another recent development is the emergence of the minimal-$\tau$ 
approximation (MTA) as a more general closure for mean-field electrodynamics 
that includes the quasilinear or first-order smoothing approximation (FOSA) as a limiting case 
(\citealt{vainshteinkitchatinov83,kleeorinetal96, rogachevskiikleeorin00, blackmanfield02}; \citetalias{bs05a}).
MTA is physically more appealing than FOSA because it takes into account the finite response time of the mean electromotive force (emf) to changes in the mean magnetic field and small-scale turbulence.
This closure leads to new terms in the mean induction equation, which becomes a telegraph-type equation \citep{couranthilbert89}, 
with second-order time derivative of the mean magnetic field.
Separate considerations, motivated in part by the need to incorporate the non-locality in the dynamo coefficients, 
lead to essentially the same telegraph-type equation \citep[][see also \citealt{hughesproctor10}]{rheinhardtb12}.
Such non-locality in time can lead to important astrophysical effects \citep[][and references therein]{hubbardb09}.
In disc galaxies, in particular, we expect memory effects to be important because
the product of the gas angular velocity $\omega$ and correlation time of the turbulence $\tau\corr$ may be of the order unity.

Yet another significant development has been the emergence of strong evidence that spiral patterns are not, 
at least in some cases, long-lived features rotating at a single constant pattern speed 
\citep{shettyetal07, dobbsetal10, sellwood11, quillenetal11, roskaretal11, wadaetal11, dobbs11, kawataetal11, khoperskovetal11}.
Theory, observations, and especially simulations of isolated and interacting galaxies point to a wide spectrum of possibilities, 
from relatively rigidly-rotating and long-lived patterns with fairly constant pattern speeds, 
to composite spirals comprised of multiple pattern speeds dominating in different radial ranges, 
to material arms that appear to rotate with the local gas velocity and thus quickly wind up.

The goal of the present work is to draw together these recent developments in dynamo theory and spiral structure
and apply the new ideas to examine non-axisymmetric regular magnetic fields in disc galaxies.
Here we focus on modes that are enslaved to the axisymmetric mode (and thus have the same growth rate in the kinematic regime); 
for a two-armed material spiral this would mean the quadrisymmetric mode (or $m=2$ mode of the Fourier expansion) that corotates with the spiral pattern, 
as well as other even-$m$ corotating modes, which are weaker.
We focus on numerical models, while
Chamandy et al.\ (2012b, hereafter Paper II) presents a semi-analytical treatment of such modes.
We leave the bisymmetric ($m=1$) mode and other odd-$m$ corotating modes to a forthcoming paper.
Where the present work differs importantly from previous work is that here we:
\begin{enumerate}
\item[(i)] incorporate MTA and explore the effects of a finite dynamo relaxation time; 
\item[(ii)] include dynamical quenching with an advective helicity flux for non-axisymmetric modes; and 
\item[(iii)] explore the effects of both steady and transient material arms on the mean magnetic field.
\end{enumerate}
In addition to this work on non-axisymmetric mean-field dynamos, 
we also briefly report on some new results from mean-field dynamo simulations which use an axisymmetric disc.

The plan of the paper is as follows. 
We present in Section~\ref{sec:equations} the basic equations and introduce our mathematical approach.
We outline the numerical model in Section~\ref{sec:numerical}.
In Sect.~\ref{sec:numerical_axisym} we discuss the findings from simulations of axisymmetric discs, 
while in Sections~\ref{sec:numerical_spiral} and \ref{sec:transient} we describe the results of simulations of non-axisymmetric discs.
A discussion of results and our conclusions are presented in Section~\ref{sec:conclusion}.

\section{The mean-field dynamo}
\label{sec:equations}

Consider the induction equation given by
\begin{equation}
\frac{\del\bm{B}}{\del t}=\bmDel\cro(\bm{U}\cro\bm{B}-\eta\bmDel\cro\bm{B}),
\end{equation}
where $\bm{B}$ is the magnetic field, $\bm{U}$ is the velocity field, $\eta$ is the magnetic diffusivity.
We follow the mean-field approach where the velocity and magnetic fields are each written as the sum of an average and a 
random
component,
\begin{equation}
\label{mean_field}
\bm{B}=\meanv{B}+\bm{b} \quad {\rm and} \quad \bm{U}=\meanv{U}+\bm{u}.
\end{equation}
Here an overbar formally represents ensemble averaging but for practical purposes can be thought of as spatial averaging over scales larger than the turbulent scale but smaller than the system size.
Substituting Eq.~(\ref{mean_field}) into the induction equation leads to the mean-field induction equation \citep{moffatt78, krauseradler80},
\begin{equation}
\label{meaninduction}
\frac{\partial \meanv{B}}{\partial t} = \bmDel \times \left( \meanv{U} \times \meanv{B} + \bgreek{\Emf} -\eta \nabla \times \meanv{B}\right).
\end{equation}
where
\begin{equation}
\label{meanemf}
\bgreek{\Emf}= \overline{\bm{u}\times\bm{b}}
\end{equation}
is the mean electromotive force.

Expressing $\Emf$ in terms of the mean field is a standard closure problem.
A relatively simple and widely used closure is the
quasilinear approximation, also known as the first-order smoothing 
approximation (FOSA; \citealt{moffatt78, krauseradler80}; \citetalias{bs05a}).
In the quasilinear theory or FOSA, one neglects nonlinear terms in the evolution equation for $\bm{u}$ and $\bm{b}$,
which, however, are retained if an evolution equation for $\Emf$ is used instead (see below).
For isotropic, helical turbulence, 
it leads to an expansion whose lowest-order terms are given by
\begin{equation}
\label{fosa}
\Emf=\alpha\meanv{B}-\eta\turb\bmDel\cro\meanv{B},
\end{equation}
where, in the kinematic limit, $\alpha=\alpha\kin$ with
\begin{equation}
\label{ttc}
\alpha\kin = -\tfrac{1}{3}\tau\corr\overline{\bfu\cdot{\bmDel\cro \bfu}}, 
\quad  
\eta\turb=\tfrac{1}{3}\tau\corr\overline{\bfu^2},
\end{equation}
and $\tau\corr$ is the correlation time of the random flow.
The turbulent transport coefficients $\alpha$ and $\eta\turb$ are proportional, 
respectively, to the mean kinetic helicity and mean energy density of 
the turbulence.
Below, we refer to the application of FOSA 
as the `standard prescription'.

An alternative treatment, suggested by 
\citet{rogachevskiikleeorin00} and \citet{blackmanfield02} was
to replace the triple correlations which arise in the evolution equation for $\Emf$ by a damping term proportional
to $\Emf$ itself (see also \citealt{vainshteinkitchatinov83,kleeorinetal96}; \citetalias{bs05a}).
Under this approximation, called the minimal-$\tau$ approximation (MTA) by \citetalias{bs05a}, one 
obtains, instead of (\ref{fosa}), an evolution equation for $\Emf$, given by
\begin{equation}
\label{minimaltau}
\frac{\partial \Emf}{\partial t}=\frac{1}{\tau\corr}(\alpha\meanv{B}-\eta\turb \bmDel\times\meanv{B})-\frac{\Emf}{\tau},
\end{equation}
where again in the kinematic limit, $\alpha$ and $\eta\turb$ are given by Eq.~\eqref{ttc} and 
$\tau$ is a relaxation time. For simplicity, 
this equation has been derived assuming that $\tau$ is scale-independent.
When one takes into account the Lorentz force, the $\alpha$-coefficient acquires an additional term
proportional to the electric current helicity (\citealt{pouquetetal76, kleeorinruzmaikin82, 
gruzinovdiamond94, blackmanfield00, radleretal03}; \citetalias{bs05a}), and then
\begin{equation}
\label{alpha_nonlinear}
\alpha = \alpha\kin + \alpha\magn = 
-\tfrac{1}{3}\tau\corr\left[\,\overline{\bfu\cdot\bmDel\cro\bfu}
- \frac{1}{4\pi\rho}\overline{\bmb\cdot\bmDel\cro \bmb}\,\right],
\end{equation}
where $\rho$ is the density.

The $\tau$-approximation is motivated by the observation that if, hypothetically, 
the mean fields were suddenly switched off then one would expect 
the mean emf to decay gradually, over a finite damping time $\tau$. 
This approximation has been tested in direct
numerical simulations of forced turbulence 
(\citetalias{bs05a}; \citealt{bs05b,bs07}). 
These simulations of MTA find that $\tau$ is positive
and the associated Strouhal number is of order unity, $\tau u\rms k\f \simeq 1$,
where $k\f$ is the wavenumber corresponding to the correlation scale of the random flow, and 
$u\rms$ is its rms velocity.
In principle,  the damping or relaxation time $\tau$ can be different
from the correlation time $\tau\corr$.
For example, if $\tau\corr$ is determined by the frequency with 
which expanding supernova shocks encounter a given point in space 
then it could be shorter than $\tau \simeq (u\rms k\f)^{-1}$ \citep{sh04}. 
Thus, we keep the ratio $c_\tau=\tau/\tau\corr$ 
as a dimensionless free parameter in the equations; for the numerical
solutions we set it to unity. We then have
\begin{equation}
\label{minimaltau2}
\left(\frac{\partial}{\partial t} +\frac{1}{\tau} \right)\Emf=\frac{c_\tau}{\tau} \left( \alpha 
\meanv{B} -\eta\turb \bmDel \cro  \meanv{B} \right).
\end{equation}
If the explicit time derivative is neglected (valid if $\Emf$ varies on timescales
long compared to the relaxation time $\tau$), 
and $\tau$ is approximated as $\tau\corr$, then (\ref{minimaltau2}) reduces to the 
expression (\ref{fosa}) obtained from the standard prescription.
An alternative way of arriving at Eq. \eqref{minimaltau2} (with $c_\tau=1$) is by keeping a time 
derivative of $\meanv{B}$ in the expression (\ref{fosa}) for $\Emf$ in order to introduce 
non-locality in time \citep{rheinhardtb12}.

We now apply the mean-field approach to the induction equation with the MTA closure.
Operating on Eq.~(\ref{meaninduction}) with $\partial/\partial t + 1/\tau$, using Eq.~(\ref{minimaltau2}), assuming $\meanv{U}$ to be independent of time, and taking $\eta$ and $\eta\turb$ to be spatially uniform, we arrive at
\[
\label{telegraphgeneral}
\begin{split}
&\left( \frac{\partial}{\partial t}+\frac{1}{\tau} \right) \frac{\partial \meanv{B}}{\partial t} = \bmDel\times \left(\meanv{U} \times \frac{\partial \meanv{B}}{\partial t} \right)+\eta \nabla^2 \frac{\partial \meanv{B}}{\partial t}\\
&\quad+\frac{1}{\tau} \left[ \bmDel \times \left(\meanv{U} \times \meanv{B}+c_\tau\alpha \meanv{B} \right)+\left(\eta + c_\tau\eta\turb \right)\nabla^2 \meanv{B}\right].
\end{split}
\]
After multiplying through by $\tau$, this leaves us with an equation containing new terms proportional to $\tau$ 
(that do not emerge under the standard prescription),
\begin{equation}
\label{telegraph}
\begin{split}
\tau \frac{\partial^2 \meanv{B}}{\partial t^2}+\frac{\partial \meanv{B}}{\partial t}&=
\tau \bmDel \times\left( \meanv{U} \times \frac{\partial \meanv{B}}{\partial t} \right)
+\tau\eta\Del^2\frac{\del\meanv{B}}{\del t}\\
&\quad+\bmDel \times \left( \meanv{U} \times \meanv{B} + c_\tau\alpha \meanv{B} \right)
+ (\eta+c_\tau\eta\turb) \nabla^2 \meanv{B}.
\end{split}
\end{equation}
The same approach applied with the standard prescription gives the familiar result,
\begin{equation}
\label{mean_induction}
\frac{\del\meanv{B}}{\del t} =\bmDel\cro(\meanv{U}\cro\meanv{B}+\alpha\meanv{B})+(\eta+\eta\turb)\Del^2\meanv{B},
\end{equation}
which is the $\tau\rightarrow0$ limit of Eq.~\eqref{telegraph} with $c_\tau=1$.
Equation~\eqref{telegraph} contains a second-order, as well as 
first-order  time derivatives, and 
belongs to the class of equations 
known as the telegraph equation \citep[e.g.,][]{couranthilbert89}. 
The second time derivative can lead to wave-like properties, with $1/\tau$ as a damping coefficient.
Unlike the mean induction equation \eqref{mean_induction}, 
the telegraph equation \eqref{telegraph} is not invariant under transformation to a rotating frame (see Paper II).

The dynamics of $\alpha\magn$ is described by the helicity evolution equation \citep{sb06},
\begin{equation}
\label{helicity}
\frac{\del\chi}{\del t}=-2\Emf\cdot\meanv{B}-2\eta\mean{\bmb\cdot\bmDel\cro\bmb}-\bmDel\cdot\Flux,
\end{equation}
where $\chi$ is the small-scale magnetic helicity density and $\Flux$ is its flux.
It is argued in \citet{sssb06} that
\begin{equation}
\label{alpm}
\alpha\magn=\frac{\eta\turb\chi}{l^2 B\eq^2}, \quad \mean{\bmb\cdot\bmDel\cro\bmb}=\frac{\chi}{l^2},
\end{equation}
where $l$ is the energy-carrying scale of the turbulence {\bf and $B\eq=\sqrt{4\pi\rho} u$ is the equipartition field strength.}
Using Eqs.~\eqref{alpm}, Eq.~\eqref{helicity} can be rewritten as an evolution equation for $\alpha\magn$,
\begin{equation}
\label{dalpha_mdt}
\frac{\del\alpha\magn}{\del t}
=-\frac{2\eta\turb}{l^2}\left(\frac{\Emf\cdot\meanv{B}}{B\eq^2}+\frac{\alpha\magn}{\Rm}\right)-\bmDel\cdot\Flux_\alpha,
\end{equation}
where $\Rm=\eta\turb/\eta$ is the magnetic Reynolds number.
Here we consider a flux of the form
\begin{equation}
\label{flux_alpha}
\Flux_\alpha=\frac{\eta\turb}{l^2B\eq^2}\Flux=\alpha\magn\meanv{U}-\kappa\bmDel\alpha\magn,
\end{equation}
where the first term is related to the advective flux of magnetic helicity \citep{sssb06}, 
while the second term leads to a Fickian diffusion of $\alpha\magn$ \citep{kleeorinetal02,betal09}. 
The latter term has been argued to exist on physical and phenomenological grounds, 
and it has been found in direct numerical simulations that $\kappa\approx0.3\eta\turb$ 
\citep[][see also \citealt{candelaresietal11}]{mitraetal10,hubbardb10}.
It is worth noting that an alternate quenching formalism has recently been suggested in the literature \citep{hubbardb11,hubbardb12},
but the question of its applicability to the galactic dynamo problem 
(with, e.g. non-periodic boundary conditions) requires further consideration which is beyond the scope of this paper.
Our aim is not to investigate different quenching formalisms,
but rather to study magnetic arms in a saturated state that is realistic.

\section{Numerical solutions}
\label{sec:numerical}
\subsection{Method and approximations used}
\label{sec:numerical_treatment}
We now examine the evolution of the mean magnetic field numerically.
We solve Eq.~\eqref{telegraph} when $\tau$ is finite, while Eq.~\eqref{mean_induction} is solved for the case $\tau\rightarrow0$.
The components of Eq.~\eqref{telegraph} in cylindrical coordinates are given in Appendix~\ref{sec:eqns_general}1. 
The saturation of the dynamo action is controlled by the helicity conservation.

For convenience we define 
\[
\Fmf\equiv \bmDel\cro\Emf,
\]
and assume that $\eta$, $\eta\turb$, $\tau$ and $c_\tau$ are constants. 
The quantity $\Fmf$ has been introduced in addition to $\Emf$ because this allows us to 
avoid having to impose boundary conditions on the components of $\Emf$ and also to 
avoid applying the no-$z$ approximation to the components of $\bmDel\times\Emf$ (see below).
Alternatively, we could have solved for the vector potential $\meanv{A}$ and $\Emf$,
so the choice of using $\meanv{B}$ and $\Fmf$ is a matter of preference.

We make the thin-disc approximation, which implies $r^{-1}\del/\del\phi\ll\del/\del z$ and $\del/\del r\ll\del/\del z$.
We also assume that the variations of the rotational and radial velocities along the $z$-axis are negligible.  
Then we can neglect $\mbz\del\mean{U}_r/\del z$ in the equation for $\mbr$ 
and $\mbz\del\mean{U}_\phi/\del z$ in the equation for $\mbp$. 
The thin-disc approximation and the condition $\bmDel\cdot\meanv{B}=0$ together imply that $\mbz/|\meanv{B}|\ll 1$. 
This approximation also allows us to ignore terms containing the $\phi$- and $r$-derivatives of 
$\alpha\mbz$ in Eqs.~\eqref{tau_r} and \eqref{tau_phi}, and of $\mbz$ in Eqs.~\eqref{Emfr} and \eqref{Emfphi}, 
as well as the term $\Emfz\mbz$ in Eq.~\eqref{dalpha_mdt} for $\alpha\magn$.
Thus, the evolution equation for $\Emfz$ is no longer needed, and all terms 
containing $\mbz$ are eliminated from all the relevant evolution equations.
(We note in passing that for solutions with quadrupolar geometry, $\mbz=0$ at the midplane, but not elsewhere.)
We can estimate the magnitude of $\mbz$ from $\bmDel\cdot\meanv{B}=0$. 
Within this framework, we do not have to worry about satisfying $\bmDel\cdot\meanv{B}=0$ and $\bmDel\cdot\Fmf=0$ numerically. 

Further, we write $\mean{U}_\phi=r\omega$, and take $\omega$ to be independent of $\phi$.
We also adopt $\mur=0$. 
However, a vertical velocity $\muz$, perhaps due to a galactic fountain flow, can be essential for the dynamo action, so it is retained.   
Note that gas outflowing from the disc can be replenished by the fountain flow from the gaseous halo 
or accretion from the intergalactic medium \citep{putmanetal12}.
The vertical flow advects magnetic field as well as magnetic helicity \citep{sssb06}.
Crucially, a magnetic helicity flux away from the dynamo region  
is required to alleviate the catastrophic quenching of the $\alpha$ effect \citepalias{bs05a}.

The no-$z$ approximation (\citetalias{sm93}; \citealt{mo95, phillips01}) is used as discussed in detail in Appendix~\ref{sec:noz}, 
to approximate the $z$-derivatives of the mean magnetic field.
This reduces the three-dimensional problem to that in two dimensions ($r$ and $\phi$), 
thus greatly reducing the computational time required.
The no-$z$ approximation provides remarkably accurate solutions 
in one-dimensional dynamical quenching models (compare \citealt{sssb06} and \citealt{sss07})
and we have also confirmed that it performs equally well in higher-dimensional problems (see Appendix~\ref{sec:noz2D}).
Under this approximation, $\muz$ represents a vertically averaged, mass-weighted vertical velocity.

\subsection{Dimensionless governing equations}
The values of the key parameters for our models are given in Table~\ref{tab:params}. 
These can be found, for example, in \citetalias{rss88,mo96}; \citet{beetal96}.
We adopt the length scale and rms velocity of the largest turbulent eddies
to be $l=100\,{\rm pc}$ and $u=10\,{\rm km\,s^{-1}}$, respectively.
These parameters are taken to be constant in space and time.
This leads to the following estimate for the turbulent diffusivity:
$\eta\turb\simeq lu/3=10^{26}\,{\rm cm^2\,s^{-1}}$.
It is convenient to use dimensionless variables, with distance along the $z$-axis  
measured in the unit of the characteristic half-thickness of the disc $h_0$, 
horizontal lengths measured in the radial disc scale length $R$, 
and time measured in typical vertical turbulent diffusion time $t\f= h_0^2/\eta\turb$. 
We take as fiducial values $h_0=500\,{\rm pc}$ and $R=20\,{\rm kpc}$.
For our purposes, it is not necessary to specify the magnetic field unit $B_0$, 
so we leave it arbitrary. 
The quantities $h_0$ and $R$ enter only
in the ratios $\lambda=h_0/R$ and $K=2(h_0/l)^2$.
The rotational shear is denoted $G=rd\omega/dr$,
$B\eq$ is a characteristic field that may vary over space and time (normally associated with the equipartition field),
and $\kappa$ is the turbulent diffusivity for $\alpha\magn$.
Then our set of equations can be written in dimensionless form as
\begin{align}
\label{ind7_r}
\frac{\del\mbr}{\del t}=&
	-\omega\frac{\del\mbr}{\del\phi}-\frac{\muz\mbr}{h} +\Fmfr,\\
\label{ind7_phi}
\frac{\del\mbp}{\del t}=& 
	G\mean{B}_r-\omega\frac{\del\mbp}{\del\phi}-\frac{\muz\mbp}{h}+\Fmfphi,
\end{align}
\begin{align}
\label{tau7_r}
\frac{\del\Fmfr}{\del t}=&
\tau^{-1}\left[
	-\frac{2c_\tau}{\pi h}\alpha\mbp 
	-c_\tau \frac{\pi^2}{4h^2}\mbr
	\right. \nonumber\\
&\left.\mbox{}	
	+c_\tau\lambda^2
		\left(
		\widehat{\mathcal{P}}\mbr
			-\frac{2}{r^2}\frac{\del\mean{B}_\phi}{\del \phi}
		\right)
	-\Fmfr
\right],\\
\label{tau7_phi}
\frac{\del\Fmfphi}{\del t}=&
\tau^{-1}\left[
-\frac{2c_\tau}{\pi h}\alpha\mbr -c_\tau\frac{\pi^2}{4h^2}\mbp
			\right.\nonumber\\
&\left.\mbox{}
	+c_\tau\lambda^2
		\left(
		\widehat{\mathcal{P}}\mbp
		+\frac{2}{r^2}\frac{\del\mean{B}_r}{\del \phi}
		\right)
	-\Fmfphi
\right],
\end{align}
\begin{align}
\label{alp_m7}
\frac{\del \alpha\magn}{\del t}=& 
-K
	\left(\frac{\bgreek{\mathcal{E}}\cdot\meanv{B}}{B\eq^2} +\Rm^{-1}\alpha\magn\right)
-\frac{\alpha\magn\muz}{h}
-\omega\frac{\del\alpha\magn}{\del\phi}\nonumber\\
&\quad+\kappa\left[\frac{\lambda^2}{r}\frac{\del}{\del r}\left(r\frac{\del\alpha\magn}{\del r}\right)+\frac{\lambda^2}{r^2}\frac{\del^2\alpha\magn}{\del\phi^2}-\frac{\pi^2}{4h^2}{\alpha\magn}\right],
\end{align}
\begin{align}
\label{Emf7_r}
\frac{\del\Emfr}{\del t}=&
\tau^{-1}\left(c_\tau\alpha\mbr-c_\tau\frac{\pi}{2h}\mbp-\Emfr\right),\\
\label{Emf7_phi}
\frac{\del\Emfphi}{\del t}=&
\tau^{-1}\left[c_\tau\alpha\mbp
+c_\tau\frac{\pi}{2h}\left(
	1+\frac{3}{4\pi}\sqrt{\frac{-D}{\pi}}
	\right)\mbr
-\Emfphi\right],
\end{align}
where 
\[
\widehat{\mathcal{P}}=
-\frac{1}{r^2} 
+\frac{1}{r}\frac{\del}{\del r} 
+\frac{\del^2}{\del r^2}
+\frac{1}{r^2}\frac{\del^2}{\del\phi^2},
\]
\[
\bgreek{\mathcal{E}}\cdot\meanv{B}=\mathcal{E}_r\mbr+\mathcal{E}_\phi\mbp.
\]
We have also neglected terms proportional to the inverse magnetic Reynolds number $\Rm^{-1}$ in Eqs.~\eqref{ind7_r} and \eqref{ind7_phi},
as $\Rm \gg 1$ in galaxies. 
We have retained such a term in 
Eq.~\eqref{alp_m7}; however, it can also be neglected if
the flux is dominant,
which is usually the case.

For $\tau\rightarrow0$, these equations reduce to the three standard equations 
of the slab dynamo:
\begin{align}
\label{ind8_r}
\frac{D\mean{B}_r}{Dt}=&
-\frac{2c_\tau}{\pi h}\alpha\mbp
-c_\tau\frac{\pi^2}{4h^2}\mbr
	\nonumber\\
&\mbox{}+c_\tau\lambda^2
	\left[
		\widehat{\mathcal{P}}\mbr
		-\frac{2}{r^2}\frac{\del\mean{B}_\phi}{\del \phi}
	\right],\\
\label{ind8_phi}
\frac{D\mean{B}_\phi}{Dt}=&
G\mean{B}_r
-\frac{2c_\tau}{\pi h}\alpha\mbr
-c_\tau\frac{\pi^2}{4h^2}\mbp
	\nonumber\\
&\mbox{}+c_\tau\lambda^2
	\left[
		\widehat{\mathcal{P}}\mbp
		+\frac{2}{r^2}\frac{\del\mean{B}_r}{\del \phi}
	\right],
\end{align}
\begin{align}
\label{alp_m8}
\frac{D\alpha\magn}{Dt}=& 
-K
	\left[
		c_\tau\alpha\left(\frac{\mbr^2+\mbp^2}{B\eq^2}\right)
		+\Rm^{-1}\alpha\magn
		\phantom{\frac{0}{0}}\right.\nonumber\\
&\left.\mbox{}		
	+c_\tau\frac{3\sqrt{-D}}{8\pi^{1/2}h B\eq^2}\frac{\mbr\mbp}{B\eq^2}
	\right]+\kappa\left[\frac{\lambda^2}{r}\frac{\del}{\del r}\left(r\frac{\del\alpha\magn}{\del r}\right)\right.\nonumber\\
&\left.\mbox{}
        +\frac{\lambda^2}{r^2}\frac{\del^2\alpha\magn}{\del\phi^2}-\frac{\pi^2}{4h^2}{\alpha\magn}\right],
\end{align}
where
\[
\frac{D}{Dt} = \frac{\del}{\del t}+\omega\frac{\del}{\del\phi}+\frac{\muz}{h}.
\]

\begin{table*}
\begin{center}
\caption{List of parameters and key dependent quantities for all models (except Model~D).
}
\label{tab:params}
\begin{tabular}{lcccccccc}
\hline
Description &Symbol &Expression &Value &Dimensionless   \\
\hline
Radial location of disc boundary       &$R$               &~                      &$20\kpc$           &$R$        \\
Disc half-thickness at $r=R/2$         &$h\f$             &~                      &$0.5\kpc$          &$h\f$        \\
Characteristic field strength at $r=0$ &$B\f$             &~                      &--                 &$B\f$       \\
Scale length of the turbulence         &$l$               &~                      &$0.1\kpc$          &$h\f/5$      \\
rms velocity of the turbulence         &$u$               &~                      &$10\kms$           &15$h\f t\f^{-1}$      \\
Turbulent diffusivity                  &$\eta\turb$       &$l u/3$                 &$10^{26}\cmcms$    &$h\f^2 t\f^{-1}$       \\
Vertical diffusion time at $r=R/2$     &$t\f$      &$h\f^2/\eta\turb$       &$0.73\Gyr$         &$t\f$       \\
Correlation time of the turbulence     &$\tau\corr$       &$l/u$                   &$10\Myr$           &$t\f/75$      \\
Disc flaring radius                    &$r\D$             &~                      &$10\kpc$           &$R/2$    \\
Disc half-thickness at $r=0$           &$h\D$             &~                      &$0.35\kpc$         &$h\f/\sqrt{2}$      \\
Brandt radius                          &$r_\omega$        &~                      &$2\kpc$            &$R/10$  \\
Angular velocity at $r=0$              &$\omega\f$        &~                      &$130\kmskpc$       &$96t\f^{-1}$      \\
Corotation radius                      &$r\corot$         &~                      &$8\kpc$            &$2R/5$      \\
Pattern speed of spiral                &$\Omega$          &$\omega(r\corot)$       &$31\kmskpc$        &$23t\f^{-1}$      \\
\hline
\end{tabular}
\end{center}
\end{table*}
\begin{table*}
\begin{center}
\caption{List of numerical models.
Resolution is given as $n_r\cro n_\phi$.
The value of $R_U$ is given at $r=R/2=10\,{\rm kpc}$.
`R' indicates a Gaussian random seed.
`T' indicates that the imposed spiral is transient, 
while `W' indicates that the $\alpha$-spiral begins as a bar and winds up with the gas.
}
\label{tab:sims}
\begin{tabular}{ccccccccccccc}
\hline
Model   &Resolution    &Seed &$l\:(\mathrm{kpc})$ &$u\:(\mathrm{{km}\,s^{-1}})$  &$R_U$ &$U\f\:(\mathrm{{km}\,s^{-1}})$   &$\kappa\eta\turb^{-1}$   &$\epsilon_\alpha$ &$\epsilon_U$ &$-k R$ &$n$        \\
\hline
A       &$100\cro60$   &0    &0.1             &10                  &0     &0       &0           &0              &0            &--      &--         \\
B       &$100\cro60$   &0    &0.1             &10                  &0.45  &0.3     &0           &0              &0            &--      &--         \\
C       &$100\cro60$   &R    &0.1             &10                  &0.45  &0.3     &0           &0              &0            &--      &--         \\
D       &$200\cro120$  &1    &0.15            &5                   &0.45  &0.3     &0           &0              &0            &--      &--         \\
E       &$200\cro120$  &0    &0.1             &10                  &0.45  &0.3     &0           &0.5            &0            &20          &2          \\
F       &$100\cro60$   &0    &0.1             &10                  &0.90  &0.6     &0           &0              &0.5          &20          &2          \\
G       &$100\cro60$   &R    &0.1             &10                  &0.45  &0.3     &0           &0.5            &0            &20          &2          \\
H       &$100\cro60$   &0    &0.1             &10                  &0     &0       &0           &0.5            &0            &20          &2          \\
I       &$200\cro120$  &0    &0.1             &10                  &0.45  &0.3     &0           &1              &0            &20          &2          \\
J       &$200\cro120$  &0    &0.1             &10                  &0.45  &0.3     &0           &0.5            &0            &8           &2          \\
K       &$200\cro120$  &0    &0.1             &10                  &0.45  &0.3     &0           &0.5            &0            &0           &2          \\
L       &$200\cro120$  &0    &0.1             &10                  &0.45  &0.3     &0           &0.5            &0            &20          &4          \\
M       &$200\cro120$  &0    &0.1             &10                  &0.45  &0.3     &0           &0.5            &0            &20,T        &2          \\
N       &$200\cro120$  &0    &0.1             &10                  &0.45  &0.3     &0           &0.5            &0            &T,W         &2          \\
O       &$200\cro120$  &0    &0.1             &10                  &0     &0       &0.3         &0.5            &0            &20          &2          \\
P       &$400\cro240$  &0    &0.1             &10                  &0.45  &0.3     &0           &0.5            &0            &20          &2          \\
\hline
\end{tabular}
\end{center}
\end{table*}
\subsection{Initial and boundary conditions}
\label{bc}
The boundary conditions are $\mbr=\mbp=0$ at $r=0$ and $r=R$,
where the former condition applies as long as we consider non-axisymmetric modes
that are localized far away from the rotation axis.
The components of $\Emf$ are evaluated for the purpose of calculating $\Emf\cdot\meanv{B}$, 
but not for calculating the components of $\bmDel\cro\Emf$.
All variables vanish at $t=0$ except for $\mbr$ and $\mbp$, whose seed values 
are chosen either to be Gaussian random fields for $\mbr$ and zero for $\mbp$, 
or to have the functional form
\[
\mbp=\frac{r}{R}\left(1-\frac{r}{R}\right)^2{\rm e}^{-r/R}(S_0+S_1\cos\phi), \qquad \mbr=-\mbp,
\]
where $S_0$ and $S_1$ are dimensionless constants that control the amplitudes 
(relative to $B_0$) of the $m=0$ and $m=1$ components of the seed field.
The solenoidality condition $\bmDel\cdot\meanv{B}=0$ can be used to obtain
the magnitude of $\mbz$.
When we refer to the $m=0$ seed, this means $S_0=1$ and $S_1=0$, while the $m=1$ seed has $S_0=0$ and $S_1=1$.

The code uses sixth-order finite differences for the spatial derivatives and a third-order 
Runge--Kutta routine for the time derivatives, using the same algorithms as the publicly
available \textsc{Pencil Code}\footnote{http://pencil-code.googlecode.com \citep{b03}}.
We have tested the code by reproducing various known results including those of \citetalias{mo96}.

\subsection{The galaxy model}
\label{sec:galaxy_model}
For the galactic rotation curve, we use the Brandt curve,
\begin{equation}
\label{Brandt}
\omega(r)=\frac{\omega_0}{\left[1+(r/r_\omega)^2\right]^{1/2}},
\end{equation}
where $\omega_0$ and $r_\omega$ are parameters, so that
with this profile, $\omega\rightarrow\const$ as $r\rightarrow0$ (solid body rotation) and $\omega\propto 1/r$ 
for $r\gg r_\omega$ (flat rotation curve).
The rotational shear rate,
\begin{equation}
\label{G}
G(r)=r\frac{d\omega}{dr}=-\omega(r)\frac{(r/r_\omega)^2}{1+(r/r_\omega)^2},
\end{equation}
tends to zero as $r\rightarrow0$ and $G=-\omega$ for $r\gg r_\omega$,
with the maximum magnitude of $2\omega_0/(3\sqrt3)$ at $r=\sqrt{2}r_\omega$.

We model the disc half-thickness as a hyperboloid \citepalias{rss88},
\begin{equation}
\label{hyperboloid}
h(r)=h\D\left[1+(r/r\D)^2\right]^{1/2},
\end{equation}
where $h\D$ is the scale height at $r=0$ and $r\D$ controls the disc flaring rate.
With this form, $h\rightarrow h_D={\rm constant}$ as $r\rightarrow0$ and $h\propto r$ for $r\gg r_D$.

We take the kinetic part of the $\alpha$-coefficient to decrease with radius according to 
F.~Krause's formula $\alpha\kin\sim l^2\omega/h$ \citepalias{rss88,bs05a}.
For the models of non-axisymmetric disc, we also impose a spiral profile on $\alpha\kin$.
For a rigidly rotating spiral, we use
\begin{equation}
\label{alpha}
\alpha\kin(r,\phi,t)=\mean{\alpha}(r)\left\{1+\epsilon_\alpha\cos[n(\phi-\Omega t)-k r]\right\},
\end{equation}
where the azimuthally averaged value of $\alpha\kin$, denoted with bar, is given by
\begin{equation}
\label{alpha_bar}
\mean{\alpha}(r)=l^2\omega(r)/h(r),
\end{equation}
$\epsilon_\alpha$ sets the degree of deviation from axial symmetry, $n$ is the number of spiral arms, 
$\Omega$ is the angular velocity of the spiral pattern (i.e. the pattern speed), 
and $k$, negative for a trailing spiral, determines how tightly the arms are wound. 
The spiral modulation of $\alpha$ could be due to a variety of different effects, 
including an increase in vorticity produced by spiral shocks \citepalias{ms91}, 
an increased turbulent velocity \citep{sh98}, 
a different coherence scale in arm and interarm regions \citep{rohdeetal99}, 
or even a new form of helicity flux \citep{vishniac12}. 
We have therefore refrained from attempting to relate $\epsilon_\alpha$ to the other parameters of the model (e.g. $l$, $u$).
Rather, for simplicity and ease of interpretation, we vary $\alpha\kin$ while holding other parameters constant along the spiral.
For a spiral winding up with the differential rotation of the gas, we take
\begin{equation}
\alpha\kin(r,\phi,t)=\mean{\alpha}(r)\left[1+\epsilon_\alpha\cos\{n[\phi+\phi_0-\omega t'\}\right],
\end{equation}
where we have included an arbitrary phase $\phi_0$ and defined $t'=t-t\on$ where $t\on$ is the time of onset of the $\alpha$-spiral. 
We have also taken $k=0$, so that the $\alpha$-spiral actually starts off as a `bar' at $t'=0$.
Wherever necessary, we also require that $\alpha\kin < u$.

The model also allows for a spiral modulation of the vertical advective velocity $\muz$ of a 
similar form:
\begin{equation}
\muz=U\f\left\{1+\epsilon_U\cos[n(\phi-\Omega t)-k r]\right\},
\end{equation}
were $U\f$ is the azimuthally averaged value of $\muz$. 

Following other authors \citepalias[e.g.][]{rss88} we define the 
turbulent magnetic Reynolds numbers to quantify the strengths
of the differential rotation, $\alpha$ effect and vertical velocity:
\[
R_\omega=Gh^2/\eta\turb, 
\quad
R_\alpha=\alpha\kin h/\eta\turb,
\quad
R_U=U\f h/\eta\turb.
\]
The local dynamo number, 
\[
D=R_\alpha R_\omega=\alpha\kin G h^3/\eta\turb^2,
\]
is a dimensionless measure of the intensity of the ($\alpha\omega$) dynamo action
at a given radius. For the mean field to grow due to the local $\alpha\omega$-dynamo action alone, 
it is required that $D<D_\crit\approx-10$ \citepalias[][see also Paper~II]{rss88}.
In the limit $r\gg r_\omega$, an axisymmetric disc has $D\simeq-9(\omega h/u)^2$.

The functional form of the characteristic magnetic field is
adopted as
\begin{equation}
\label{Beq}
B\eq=B_0\Exp{-r/R};
\end{equation}
if appropriate, this can be identified with the magnetic field strength corresponding to 
the energy equipartition with the turbulence. Nonlinear dynamo effects are expected to become pronounced as soon as $|\meanv{B}|\simeq B\eq$.

\subsection{Models explored and the representation of the results}
\label{sec:models}
\begin{figure}
\includegraphics[width=84mm]{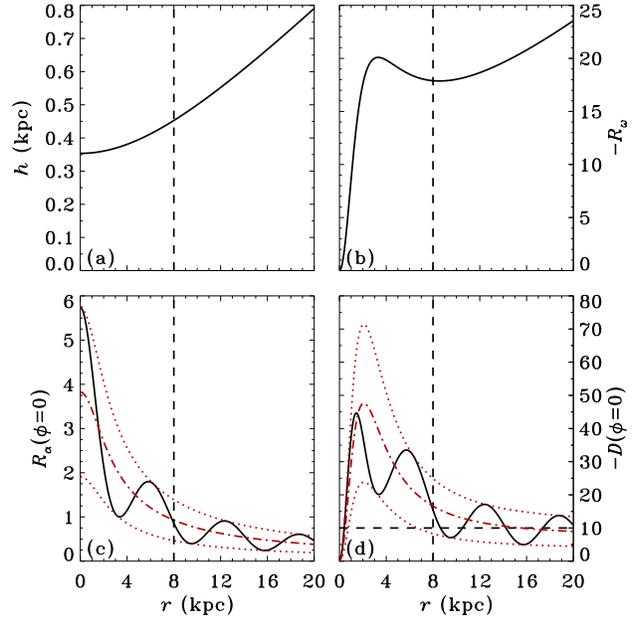}
\caption{Inputs used for the basic non-axisymmetric Model~E. 
{\bf(a)} Disc scale height $h$, given by Eq.~\eqref{hyperboloid}. 
{\bf(b)} The dimensionless quantity $-R_\omega=Gh^2/\eta\turb$ with $G$ given by Eq.~\eqref{G}. 
{\bf(c)} The dimensionless quantity $R_\alpha=\alp\kin h/\eta\turb$, with $\alp\kin$ given by Eq.~\eqref{alpha}, at $\phi=0$ (solid), azimuthal maximum/minimum (dotted), and azimuthal mean (dash-dotted). 
{\bf(d)} Negative of the dynamo number $-D=-R_\alp R_\omega$ at $\phi=0$ (solid), along with its azimuthal extrema (dotted) and mean (dash-dotted). 
The azimuthal mean is used in axisymmetric models.
Negative of the approximate critical dynamo number $-D_\crit\simeq10$ is shown dashed for reference.
(Online versions of all the figures are in colour.)
\label{fig:inputs}}
\end{figure}

The various numerical models considered here are summarized in Table~\ref{tab:sims}.
We adopt $r_\omega=0.1R=2\,{\rm kpc}$ and set the circular speed to 
$\mup=375h_0t\f^{-1}=250\,{\rm km\,s^{-1}}$ at $r=R/2=10\,{\rm kpc}$, which gives 
$\omega_0=96t\f^{-1}=130\,{\rm km\,s^{-1}\,kpc^{-1}}$.
This implies that 
$\omega=18.75t\f^{-1}=25\,{\rm km\,s^{-1}\,kpc^{-1}}$, $G\simeq-18t\f^{-1}=-24\,{\rm km\,s^{-1}\,kpc^{-1}}$, 
$\overline{\alpha}=0.75h_0t\f^{-1}=0.5\,{\rm km\,s^{-1}}$ and the azimuthally averaged dynamo number $\overline{D}\simeq-13.5$ 
at $r=R/2=10\,{\rm kpc}$.
The resulting radial profiles of $h$, $R_\omega$, $R_\alpha$, and $D$ are shown in 
Fig.~\ref{fig:inputs}.

The corotation radius for models with steady spiral forcing is chosen as $r\corot=8\kpc$, which is close to the observational estimates for the Milky Way \citep{gerhard11,acharovaetal11}.
Moreover, a reasonable estimate for $\tau$ is given by $\tau=l/u=(l^2/3h_0^2)t\f=t\f/75$.
It is possible that $\tau$ is much smaller than our simple estimate for some galaxies and much larger for others.
Therefore, we also consider the $\tau\rightarrow0$ case, as well as $\tau=2l/u$ for some models.

We decompose each component of the mean magnetic field $\meanv{B}$ into a 
cosine Fourier series, with certain phases $\phi_{0,i}^{(m)}$ (see below),
\begin{equation}
\label{Fourier}
\mean{B}_i(r,\phi,t)=\displaystyle\sum_{m=0}^\infty\widetilde{B}_i^{(m)}(r,t)
\cos\{m[(\phi-\phi_{0,i}^{(m)}(r,t)]\},
\end{equation}
where $i=r, \phi$,
Thus, 
\begin{equation}
\label{B0}
\widetilde{B}_i^{(0)}=\frac{1}{2\pi}\displaystyle\int_0^{2\pi}\mean{B}_i(r,\phi,t)\,d\phi,
\end{equation}
and, for $m>0$,
\begin{equation}
\begin{split}
\label{Bm}
\widetilde{B}_i^{(m)}=\frac{1}{\pi}\displaystyle\int_0^{2\pi}\mean{B}_i(r,\phi,t)
\cos\left\{m\left[\phi-\phi_{0,i}^{(m)}(r,t)\right]\right\} \,d\phi.
\end{split}
\end{equation}
The phase $\phi_{0,i}^{(m)}$ is obtained by trying all possible values
and choosing the one which maximizes $\widetilde{B}_i^{(m)}$.
Using their phases, 
we also determine the rotation rates of the various Fourier modes by numerically differentiating the phase with respect to time.

The azimuthal average of the magnetic energy
can be written as
\begin{equation}
\overline{E}(r,t)=\displaystyle\sum_{m=0}^\infty \widetilde{E}^{(m)}(r,t),
\end{equation}
where
\begin{equation}
\widetilde{E}^{(0)}(r,t)=\frac{1}{8\pi}\left\{
\left[\widetilde{B}_r^{(0)}(r,t)\right]^2+\left[\widetilde{B}_\phi^{(0)}(r,t)\right]^2\right\},
\end{equation}
and, for $m>0$,
\begin{equation}
\label{Em}
\widetilde{E}^{(m)}(r,t)=\frac{1}{16\pi}\left\{
\left[\widetilde{B}_r^{(m)}(r,t)\right]^2+\left[\widetilde{B}_\phi^{(m)}(r,t)\right]^2
\right\},
\end{equation}
where the additional factor $1/2$ in the latter equation arises from averaging 
$\cos^2[m(\phi-\phi_{0,i}^{(m)})]$ over the interval $(0,2\pi)$.
Averaging over the area of the disc, 
we obtain the average normalized magnetic energy in mode $m$,
\begin{equation}
\left\langle \frac{\widetilde{E}^{(m)}(t)}{B\eq^2}\right\rangle=\frac{2}{R^2}\displaystyle\int_0^R  \frac{\widetilde{E}^{(m)}(r,t)}{B\eq^2(r)}\,r\,dr.
\end{equation}

\section{Dynamo in an axisymmetric disc}
\label{sec:numerical_axisym}
A number of interesting questions can be addressed by considering dynamo action in an axisymmetric disc.
One of them is how a whole young galaxy becomes magnetized with a coherent field
given that the radial diffusion time across a galaxy, 
$t_R \simeq R^2/\eta\turb\simeq 3\times 10^2\,{\rm Gyr}$, is typically larger than the age of the Universe.
We examine this in the context of the dynamical quenching model
of the galactic dynamo.

\subsection{Axisymmetric solutions}
\label{sec:flux}
\begin{figure}
\begin{center}
\includegraphics[width=84mm]{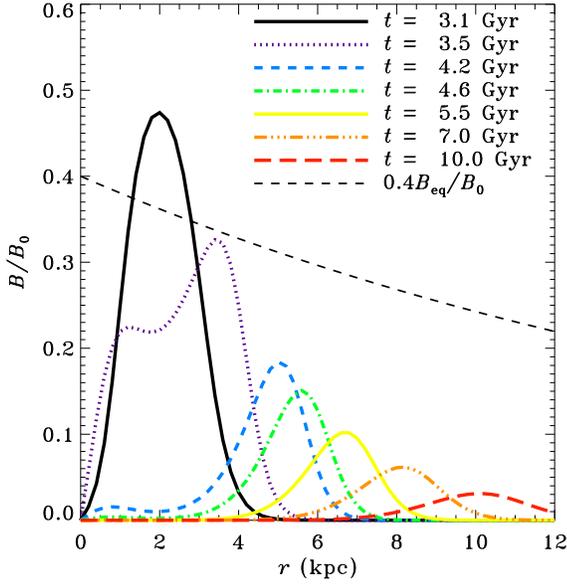}
\caption{Evolution of the field strength $B=(B_r^2+B_\phi^2)^{1/2}$ in Model~A (axisymmetric disc, with $U\f=0$, $\kappa=0$).
\label{fig:t20_dynq_190_colour_tau0}}
\end{center}
\end{figure}

\begin{figure}
\begin{center}
\includegraphics[width=84mm]{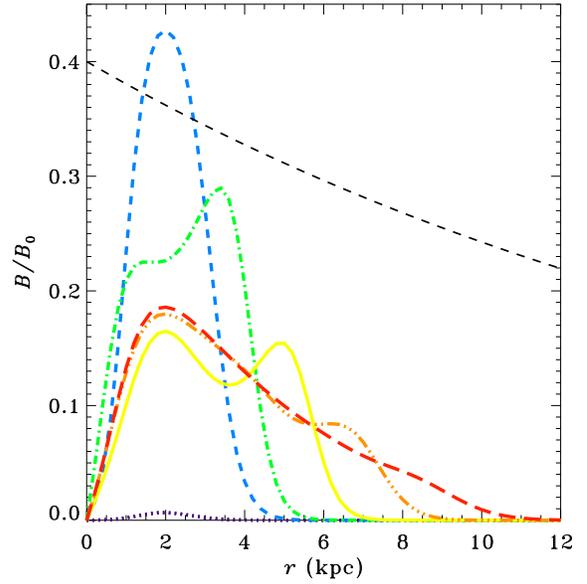}
\caption{Evolution of the field strength in Model~B (axisymmetric disc, with $U\f=0.3\kms$, $\kappa=0$).
Times plotted are as in Fig.~\ref{fig:t20_dynq_190_colour_tau0}, 
with $t=3.1\Gyr$ omitted since $\mean{B}/B\f$ is too small to be visible.
\label{fig:t20_dynq_189_colour_tau0}}
\end{center}
\end{figure}
In local dynamo models the presence of magnetic helicity fluxes have been shown to alleviate the catastrophic quenching of the dynamo
\citep{kleeorinetal02,sssb06}.
In the absence of any helicity flux,
the mean magnetic field does grow to a fraction of the equipartition value, 
until $\alpha\magn$ cancels $\alpha\kin$, after which magnetic field decays. 
The result which obtains when the radial dimension is included
is more interesting, 
even in the absence of a helicity flux.
The evolution of the radial profile of the mean field strength for such a case (Model A) 
is shown in Fig.~\ref{fig:t20_dynq_190_colour_tau0} adopting $\tau\rightarrow0$.
No notable differences exist between the $\tau\rightarrow0$ and $\tau=l/u$ cases; 
for example, the kinematic global growth rate $\Gamma=5.7 t\f^{-1}\simeq7.8\,{\rm Gyr^{-1}}$ is virtually the same.\footnote{That 
there is not much difference between the two cases is expected because in an axisymmetric disc, axisymmetric modes are dominant, 
and effects of a finite relaxation time in the axisymmetric problem are of order $\Gamma\tau\ll1$; 
only in the non-axisymmetric case do effects of order $\Omega\tau\la1$ play a role in our model.}
Since the local growth rate of the field depends on the dynamo number, 
which in turn depends on $r$, the field maximum travels in radius.
More specifically, the maximum is first localised where the dynamo number is the 
largest ($r_M$, say), but, at later time, at radii $r_M-\Delta r_1$ and $r_M+\Delta r_2$, etc. 
This leads to two rings of enhanced mean field, one moving inward and the other moving outward (the outer ring is more prominent in Model~A). Because of the catastrophic $\alpha$-quenching,
the mean field eventually becomes negligible.
(See \citet{moshsokoloff98} for a discussion of propagating magnetic fronts in disc galaxies.)

On the other hand, if a vertical advective flux or diffusive flux is present, 
catastrophic quenching is averted. 
Then magnetic field can persist at about the same strength in a wide radial range,
gradually spreading out until it occupies the entire region of the disc where the local dynamo number is supercritical.
This is the situation illustrated with Model B, as shown in 
Fig.~\ref{fig:t20_dynq_189_colour_tau0} ($\tau\rightarrow0$ case), for $R_U=0.45$ 
at $r=R/2=10\,{\rm kpc}$ and $\muz=0.3\,{\rm km\,s^{-1}}$ at all $r$.
In this model, both $\tau\rightarrow0$ and $\tau=l/u$ cases are again very similar, 
and the kinematic global growth rate for both cases is 
$\Gamma=5.1 t\f^{-1}\simeq7.0\,{\rm Gyr^{-1}}$.
This is slightly smaller than for Model~A, 
since the vertical flux removes large-scale magnetic field, rendering the dynamo less efficient in the kinematic stage.
The results are also similar if we replace the advective flux of $\alpha\magn$ with a turbulent diffusive flux.
This model clarifies  how the whole disc becomes magnetized. 
It does so by first reaching significant strengths in regions where the dynamo number is largest.
Growth then saturates at each radius at a fraction of the local equipartition value, 
provided the helicity flux is favourable for the dynamo action. 
The role of the radial diffusion is merely to couple the dynamo action at different radii
as to lead to a magnetic structure growing at a single rate. 
Importantly, the entire disc becomes magnetized over a timescale much shorter than 
the radial diffusion time.
\begin{figure}
\begin{center}
\includegraphics[width=84mm]{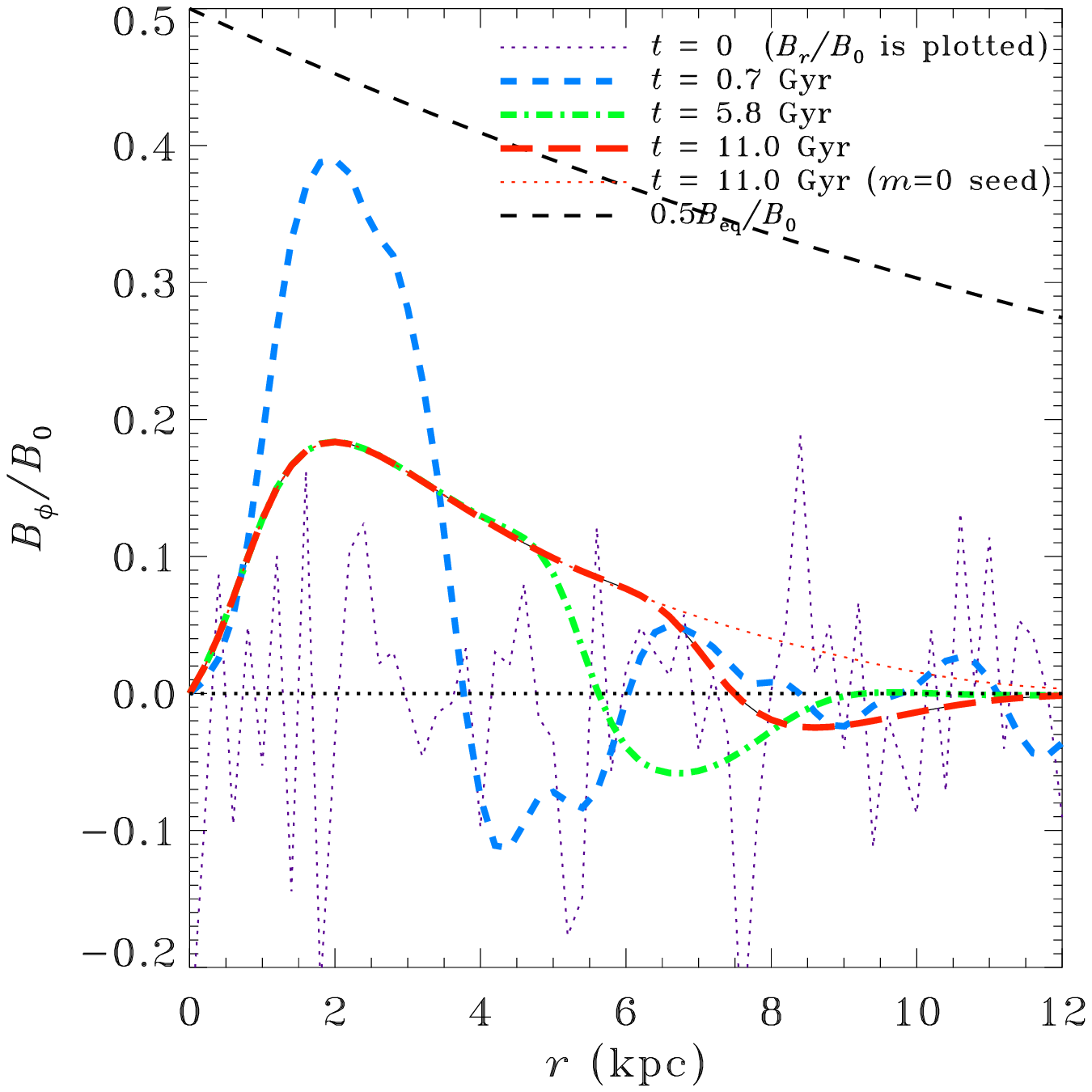}
\caption{
Evolution of the magnetic field that starts as a random seed magnetic field (Model~C) in the $\tau\rightarrow0$ case. 
Curves show $\mbp$ at the azimuth $\phi=0$ at various times, with the exception of the field at $t=0$, where $\mbr$ is shown (since $\mbp=0$).
Otherwise, $\mbr$ is omitted for the sake of clarity, but vanishes at the same set of radii as $\mbp$, 
indicating reversals of the field at these locations.
For the final time, the solution obtained with an $m=0$ seed is shown as a thin dotted line for reference.
\label{fig:rand_dynq_sf}
}
\end{center}
\end{figure}
\begin{figure*}
\begin{center}
\includegraphics[width=175mm]{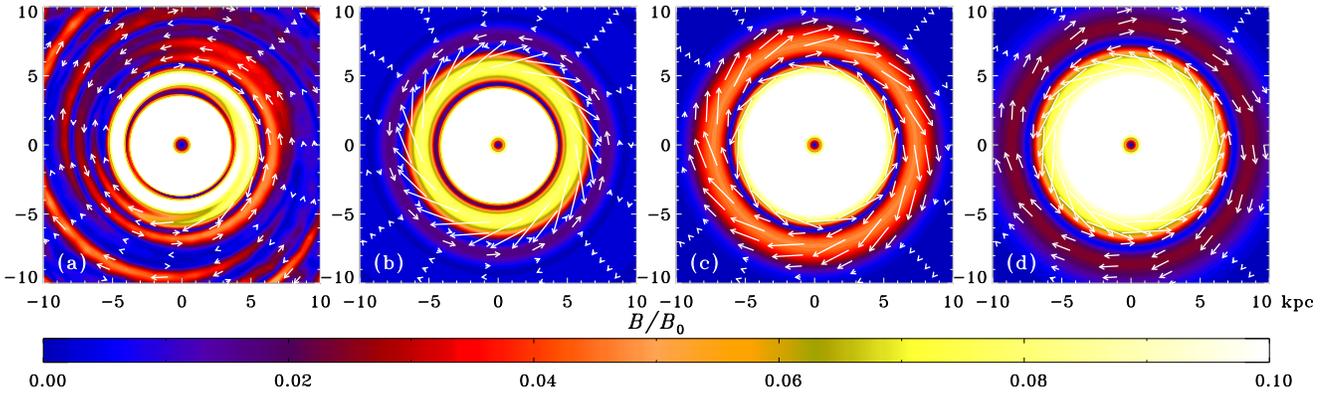}	
\caption{
The strength of the mean magnetic field (colour coded) at {\bf(a)} $t=0.7\Gyr$, {\bf(b)} $2.8\Gyr$, {\bf(c)} $7.3\Gyr$, and {\bf(d)} $11.0\Gyr$ 
in Model~C (random seed field) for the case $\tau\rightarrow0$.
The colour of the central region has been saturated to enhance visibility of the magnetic structure in the outer disc.
Field vectors are also shown, with tail length proportional to the magnitude of magnetic field, for $r>5\kpc$ only, to avoid clutter.
Reversals which evolve with time are visible (for example, the central dark ring which moves out with time).
\label{fig:rand_dynq_vec_arrow}
}
\end{center}
\end{figure*}
\subsection{Reversals of the magnetic field}
We find that if the seed magnetic field is weak enough, the steady-state magnetic configuration is independent of its form and strength. 
However, a relatively strong initial magnetic field can affect the steady-state magnetic configuration 
if nonlinear dynamo effects become important before the leading dynamo eigenfunction 
(normally represented by an axisymmetric magnetic field without any reversals along the radius) 
can become dominant \citep{sh05,moetal12}. 
In particular, a random seed magnetic field can lead to long-lived reversals in the (quasi-)steady state magnetic configuration, 
either global \citep{pss93} or localised in both radius and azimuth \citep{bpss97}. 
A suitable random seed magnetic field can be readily provided by the fluctuation dynamo action \citep{pss93}. 
Earlier results in this area have been obtained with a heuristic algebraic nonlinearity in the mean-field dynamo equation. 
Here we revisit this idea, but now with the physically motivated dynamic nonlinearity and finite $\tau$.

In Model~C, we have chosen a random seed specified as a 
two-dimensional Gaussian random field
with the root-mean-square value of about one tenth the equipartition value 
(i.e., a fraction of $\mu$G in the Solar neighbourhood).\footnote{The 
solenoidality of the seed field is ensured by an appropriate choice of $\mbz$.}
We also tried ten times larger seeds and found almost identical results.
The number of radial grid points in this models is $n_r=100$, corresponding to a resolution $0.2\,{\rm kpc}$ in radius, 
comparable to the correlation scale of the interstellar turbulence.
The results are shown in Fig.~\ref{fig:rand_dynq_sf}. 
Firstly, large scale fields can develop over kpc scale regions, even in the outer disc, on Gyr timescales.
More interestingly,
it can be seen that
reversals develop and persist for several Gyr in the nonlinear regime.
(Because of the random nature of the problem, 
other random seeds of the same strength or different resolutions can result in field configurations without reversals.)
The reversals are also apparent in a time sequence of the magnetic field of Model~C shown in Fig.~\ref{fig:rand_dynq_vec_arrow}.
At least one reversal in the regular magnetic field has been observed in the Milky Way \citep{vanecketal11}.
It can be seen from both figures that 
as time goes on, the reversals become global in nature, propagate outward on a timescale of several Gyr, and in the process decrease in number.
By $t=6t\f=4.4\Gyr$ all reversals inside $r=16\kpc$ are global in nature (i.e. occur at all azimuth for a given radius).
In the $\tau=l/u$ case, the noise from the random seed takes much longer to dissipate, 
and global reversals are only apparent much later on, at large radii where the field is very weak.
\citet{pss93} found that the persistence of the reversals at the galactic lifetime scale strongly depends on the rotation curve. 
Our experiments reported here suggest that memory effects (finite $\tau$) are also important.
\subsection{Non-axisymmetric magnetic fields in an axisymmetric disc}
\label{axisym_bisym}
\begin{figure}
\begin{center}
\includegraphics[width=84mm]{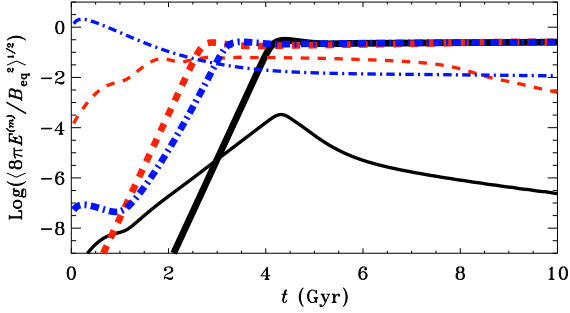}	
\caption{
Evolution of the normalised magnetic field strength averaged over the disc,
in the $m=0$ (thick) and $m=1$ (thin) modes, 
for a purely bisymmetric ($m=1$) seed magnetic field (Model~D) with $\tau\rightarrow0$.
Different line styles show three cases with identical parameters, except for the magnitude of the 
seed field, so that the dynamo action saturates 
(i)~when the $m=0$ mode dominates (solid), 
(ii)~when $m=1$ dominates (dashed), and
(iii)~right from $t=0$ (dash-dotted).
\label{fig:bisym_dynq_Gam}
}
\end{center}
\end{figure}
We find that the non-axisymmetric modes decay in in an axisymmetric disc for the 
parameters used in most of our numerical models (see Table~\ref{tab:params}).
However, the $m=1$ mode (easiest non-axisymmetric mode to excite) can grow for somewhat modified parameters
(Model~D of Table~\ref{tab:sims}), 
with $\alpha\kin$ truncated in the central region to $u/2$.

In Model~D, we focus on the case $\tau\rightarrow0$. 
The results of runs with $\tau\ne0$ do not show notable differences.
The $m=1$ mode is readily excited with the kinematic growth rate $\Gamma_1=1.7t\f^{-1}=3.5\Gyr^{-1}$. 
The growth rate for the axisymmetric mode for these parameters is $\Gamma_0=4.7t\f^{-1}=9.6\Gyr^{-1}$.
Since $\Gamma_0>\Gamma_1$ with a significant margin, 
magnetic field in a mature dynamo hosted by an axisymmetric disc will be axisymmetric 
unless the seed magnetic field is strongly non-axisymmetric. 
The latter is, in fact, quite plausible if the seed field
arises, for example, from a putative intergalactic field captured as the disc galaxy forms.
In order to address this point, we have run Model~D, starting from a 
purely $m=1$ seed, with the $m=0$ mode being seeded only by the numerical noise, as in \citetalias{mo96}.
We adjust the strength of the seed such that nonlinear dynamo effects become significant
at various important stages: (i)~when the $m=0$ mode has already come to dominate, 
(ii)~when it is still weaker than the $m=1$ mode, and (iii)~right from $t=0$.

The results are shown in Fig.~\ref{fig:bisym_dynq_Gam}. 
Thick lines correspond to the normalised magnetic field strength (averaged over the area of the disc) 
in the $m=0$ mode while thin lines correspond to that in the $m=1$ mode.
In Case~(i), shown by the solid lines, the saturation of the (essentially axisymmetric) field clearly causes the $m=1$ mode to decay since its growth requires a stronger (unquenched) dynamo action.
In Case~(ii) (dashed lines), 
the early saturation of the (stronger) $m=1$ mode still does not prevent the $m=0$ mode from growing and evolving as in Case~(i).
This happens because the magnetic field in between the extrema of the non-axisymmetric magnetic 
field remains relatively weak when the $m=1$ mode ceases to grow.
Therefore, the growing $m=0$ mode is supported by the local dynamo action in those regions,
to fill the gaps until the field is saturated everywhere to become nearly axisymmetric. 
However, the $m=1$ mode, though subdominant, does not immediately decay but remains strong for several Gyr. 
This is because the outward spreading of the dominant $m=0$ mode is preceded by the outward spreading of the $m=1$ mode.
Eventually, the axisymmetric mode comes to dominate everywhere and the $m=1$ decays more quickly (for $t\gtrsim8\,{\rm Gyr}$).
In Case~(iii) (dash-dotted lines), we find that the $m=1$ mode first decays and then remains much weaker than the 
$m=0$ mode.

As far as the overall survival of non-axisymmetric modes in an 
\textit{axisymmetric\/} disc is concerned, we can conclude that (i) a rather special parameter combination is needed for a non-axisymmetric mode to grow;
(ii) for a non-axisymmetric mode to persist for several Gyr in the nonlinear stage, 
it must reach the saturation strength before the $m=0$ mode can do so; 
given that non-axisymmetric modes have smaller growth rates, 
this implies that they must be much stronger initially;
(iii) in any case, all non-axisymmetric modes eventually decay in an axisymmetric disc.
These conclusions make it all the more reasonable to suggest that deviations from perfect
axial symmetry of the underlying disc are 
required to explain the prevalence of non-axisymmetric regular fields in many galaxies.
It is the mechanism of `spiral forcing' of the dynamo to which we now turn.
\begin{figure}
\begin{center}
\includegraphics[width=84mm]{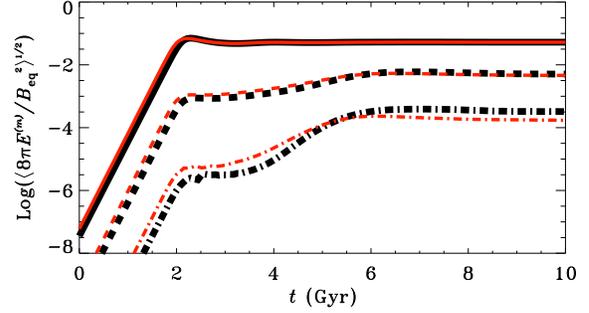}		
\caption{
Magnetic field strength in the $m=0$ (solid), $m=2$ (dashed) and $m=4$ (dash-dotted) modes of the mean magnetic field, 
normalized to the equipartition field strength $B\eq$ and averaged over the entire disc, for Model~E.
Results obtained for $\tau=l/u$ are shown in thick black, while those for $\tau\rightarrow0$ are in thin red.
All modes have exponential kinematic growth rate $\Gamma^{(m)}=7.0\,{\rm Gyr^{-1}}$.
For the convenience of presentation, time has been rescaled so that the simulation in fact starts 
from $t=-8\,{\rm Gyr}$.
\label{fig:Gam}}
\end{center}
\end{figure}
\begin{figure}
\begin{center}
\includegraphics[width=84mm]{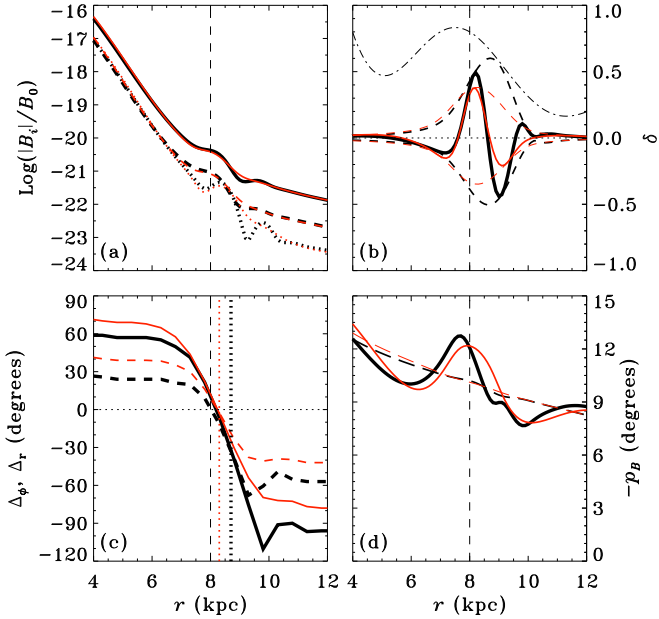}
\caption{ 
Properties of the field in the kinematic regime for Model~E at a time $3t\f$ after the simulation is begun.
Thick black illustrates the $\tau=l/u$ case while thin red illustrates $\tau\rightarrow0$.
In all plots, the corotation radius is shown as a dashed vertical line.
{\bf (a)}~The normalized azimuthal (solid), radial (dashed), and vertical (dotted) components 
of the field at the azimuth $\phi=\phi\corot$ for which the $\alpha$-spiral crosses corotation.
{\bf (b)}~The ratio $\delta$ of the non-axisymmetric to axisymmetric $\phi$-component of the field at $\phi\corot$ (solid), and at azimuthal extrema (dashed).
The quantity $\alpha\kin(r, \phi\corot)$ is shown as a thin dash-dotted line for reference.
{\bf (c)}~The phase of the $\phi$-component (solid) or $r$-component (dashed) of the magnetic spiral minus the phase of the $\alpha$-spiral (smoothed over 5 radial grid points, or 0.5\kpc).
The radius $r\ma$ at which 
the global maximum of $\delta$ occurs
is plotted as a vertical dotted line for each $\tau$ case.
{\bf (d)}~Negative of the pitch angle at $\phi=\phi\corot$ (solid), with azimuthal mean (dashed). 
\label{fig:sf_kin}}
\end{center}
\end{figure}
\begin{figure}
\begin{center}
\includegraphics[width=84mm]{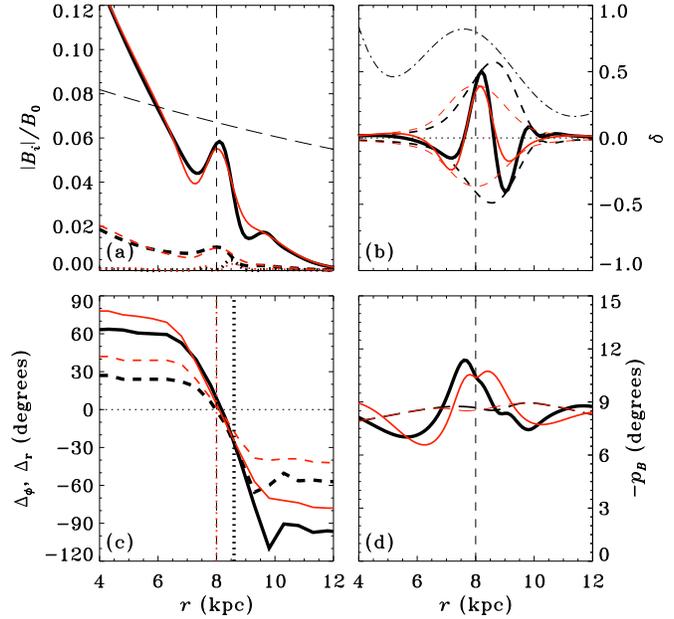}
\caption{
Same as Fig.~\ref{fig:sf_kin} (Model~E), but now for $t=10\,{\rm Gyr}$ (see Fig.~\ref{fig:Gam}), 
when the field has reached equilibrium as a result of dynamical quenching.
Note that in (a) The scale is now linear instead of logarithmic.
The long dashed curve is equal to $0.1B\eq/B_0$, shown for reference.
\label{fig:sf_dynq}}
\end{center}
\end{figure}
\section{Magnetic arms enslaved to a stationary spiral pattern}
\label{sec:numerical_spiral}

In this section, we consider  the evolution of the mean magnetic field
in a non-axisymmetric disc where the $\alpha$ effect is modulated by a stationary spiral pattern.
As above, our analysis includes the dynamic nonlinearity,
thus extending fully into the saturated states of the dynamo action, 
and allows for a finite dynamo relaxation time $\tau$.

As a consistency check, we have first made sure that the standard dynamo solutions,
obtained by solving Eq.~\eqref{mean_induction} are obtained when solving Eq.~\eqref{telegraph} with $\tau$ approaching zero.
Indeed, we found good agreement with them for $\tau=10^{-3}$.
To ensure that the results are not sensitive to the location of the outer boundary, 
we compared solutions with
outer radius of $R=15$\,kpc and 30\,kpc, with the standard value
$R=20$\,kpc (with the radial resolution modified proportionately).
The results are not sensitive to this adjustment.
 
\begin{figure*}
\begin{center}
\includegraphics[width=132mm]{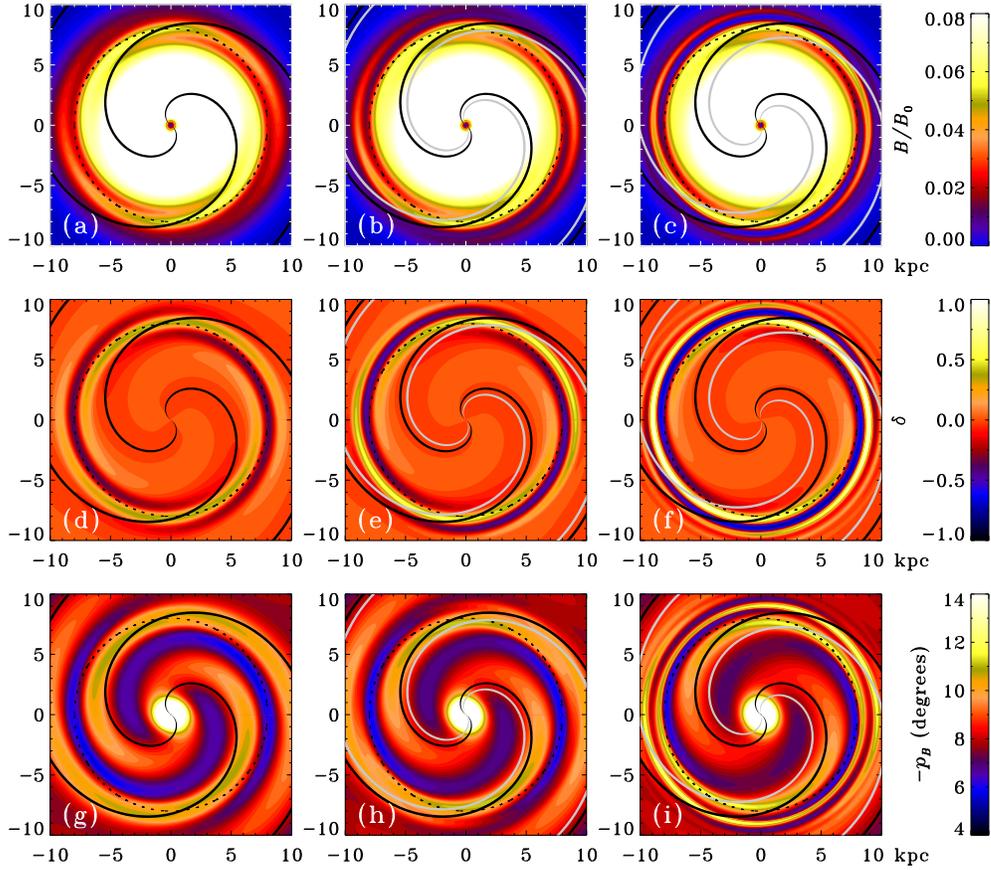}	
\caption{
The top panels show the magnitude of the magnetic field in the saturated state 
in the $\tau\rightarrow0$ (left), $\tau=l/u$ (middle) and $\tau=2l/u$ (right) cases.
In all panels, the peak of $\alpha\kin$ is shown as a thick black line,
the grey shifted line illustrates (for reference) the $\alpha\kin$ spiral shifted by the angle $-\Omega\tau$,
while the corotation radius is illustrated as a black dotted circle.
The colour of the central regions in a--c has been saturated in order to visually bring out the behaviour near corotation,
where the non-axisymmetric modes are important.
Panels d--f show the ratio $\delta$ of the non-axisymmetric to axisymmetric components of $\mbp$.
The bottom panels show the magnetic pitch angle.
Again, the colour within a region around the center has been saturated.
\label{fig:vec_dynq}}
\end{center}
\end{figure*}
\begin{figure}
$
\begin{array}{c}
  \includegraphics[width=84mm]{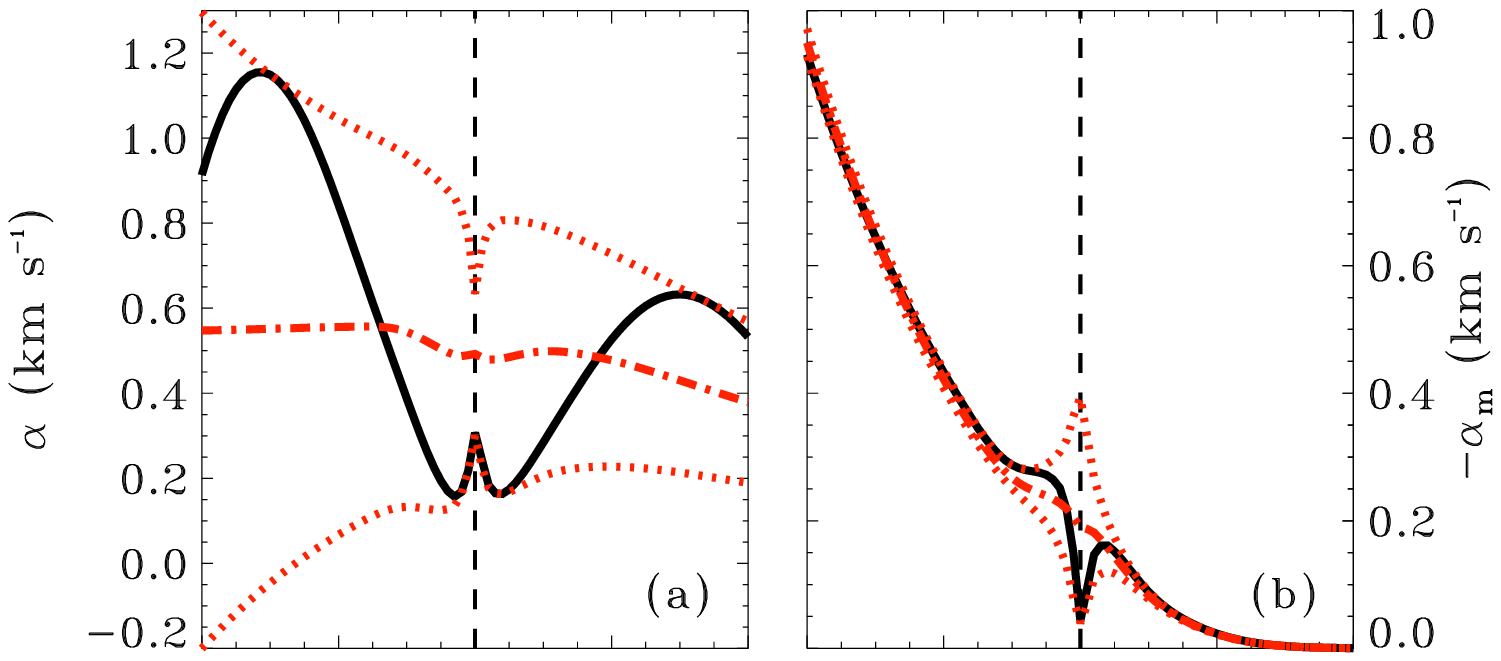}\\
  \includegraphics[width=84mm]{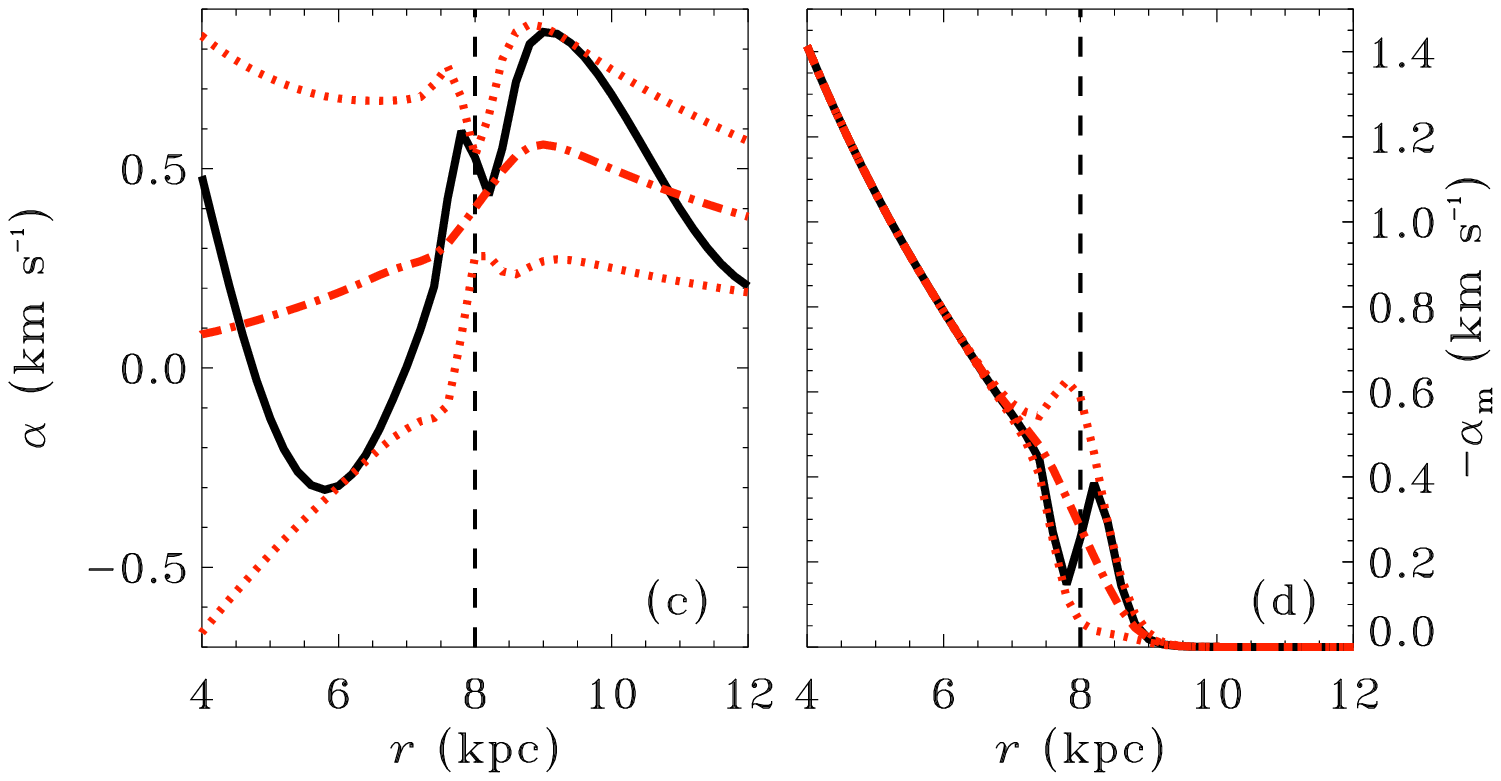}
\end{array}
$
\caption{{\bf (a)} The profile of $\alpha$ for Model~E ($\tau\rightarrow0$ case), in the saturated state, for $\phi=0$ (solid), 
azimuthal mean (dash-dotted) and envelope for all azimuth (dotted).
{\bf (b)} Same as (a) but now for $-\alpha\magn$. 
{\bf (c)} Same as (a) but now for Model~H, at the time $t=4.4\Gyr$, when the outgoing ring is passing the corotation circle.
{\bf (d)} Same as (c) but now for $-\alpha\magn$.
\label{fig:alpha_m}}
\end{figure}
Our standard model is labelled E, with parameters given in Table~\ref{tab:sims}. To explore
the parameter space, we have also considered Models~F--L.

The evolution of the normalized magnetic field strength in various even-$m$ modes, 
averaged over the area of the disc, is shown, for Model~E, in Fig.~\ref{fig:Gam}.
In the kinematic regime, all modes have about the same growth rate 
$\Gamma\simeq5.1t\f^{-1}\simeq7.0\,{\rm Gyr^{-1}}$ for both $\tau\rightarrow0$ and $\tau=l/u$ cases.
This growth rate is also in close agreement with the growth rate for the axisymmetric case (Model~B). 
This is not surprising because the average field in the disc is used to calculate the growth rate,
and the field is dominated by the axisymmetric field at $r\ll r\corot$.
The even-$m$ modes grow together with the $m=0$ mode because they are driven by (enslaved to) it.
The $m=2$ and $m=4$ modes corotate with the spiral pattern. 
\begin{figure*}
\begin{center}
\includegraphics[width=132mm]{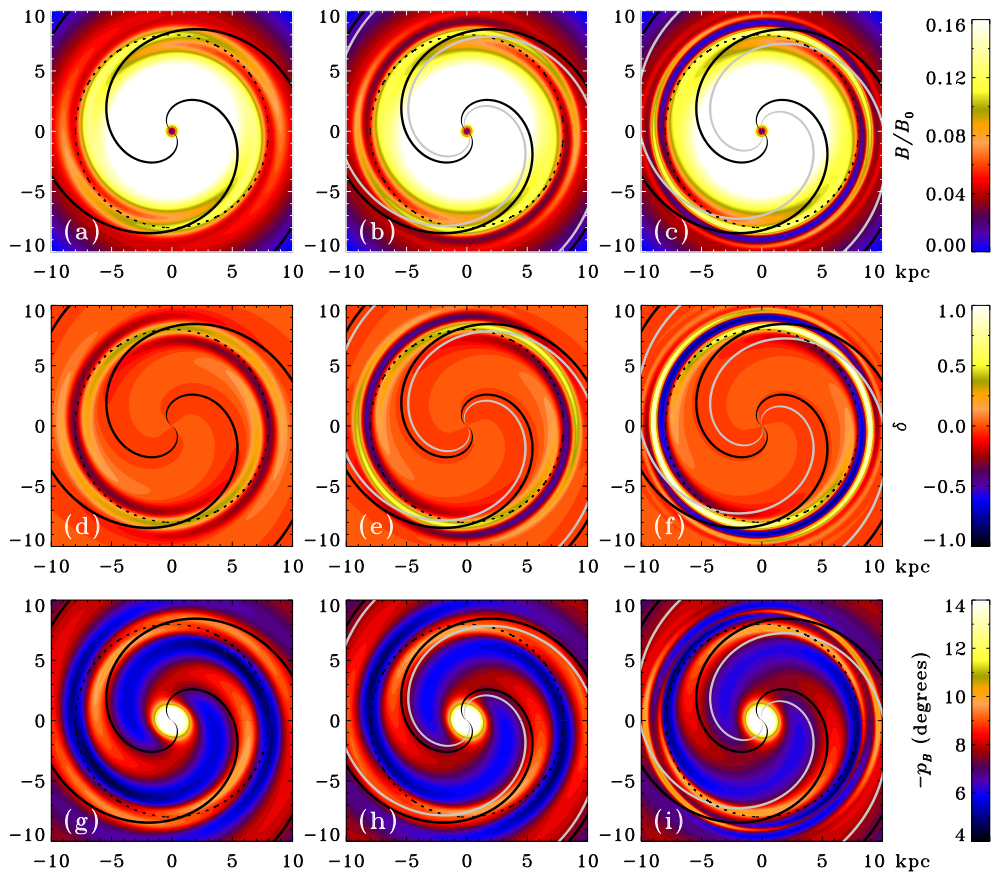}	
\caption{
Same as Fig.~\ref{fig:vec_dynq} but now for the steady state solution of Model~O (diffusive flux of $\alpha\magn$ instead of advective flux).
Note the change in plotting range in the top row.
\label{fig:vec_dynq_diffusive}}
\end{center}
\end{figure*}
\begin{figure}
\begin{center}
\includegraphics[width=84mm]{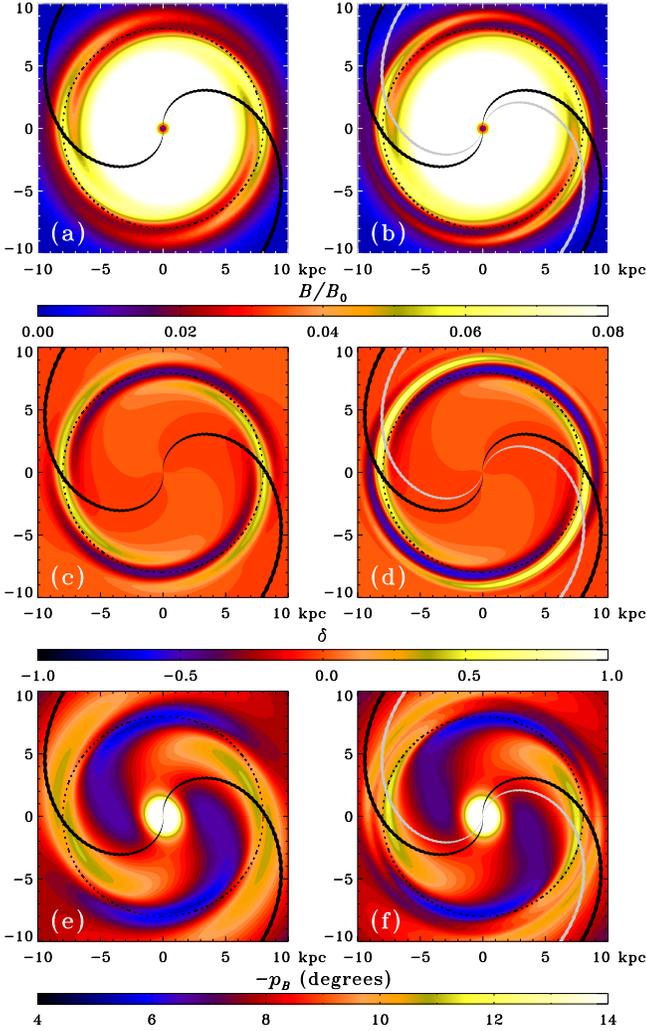}	
\caption{
Similar to Fig.~\ref{fig:vec_dynq}, but now for Model~J ($k=-8R^{-1}$ instead of $-20R^{-1}$), 
with $\tau\rightarrow0$ on the left and $\tau=l/u$ on the right.
\label{fig:kappa8_vec_dynq}}
\end{center}
\end{figure}
\begin{figure}
\begin{center}
\includegraphics[width=84mm]{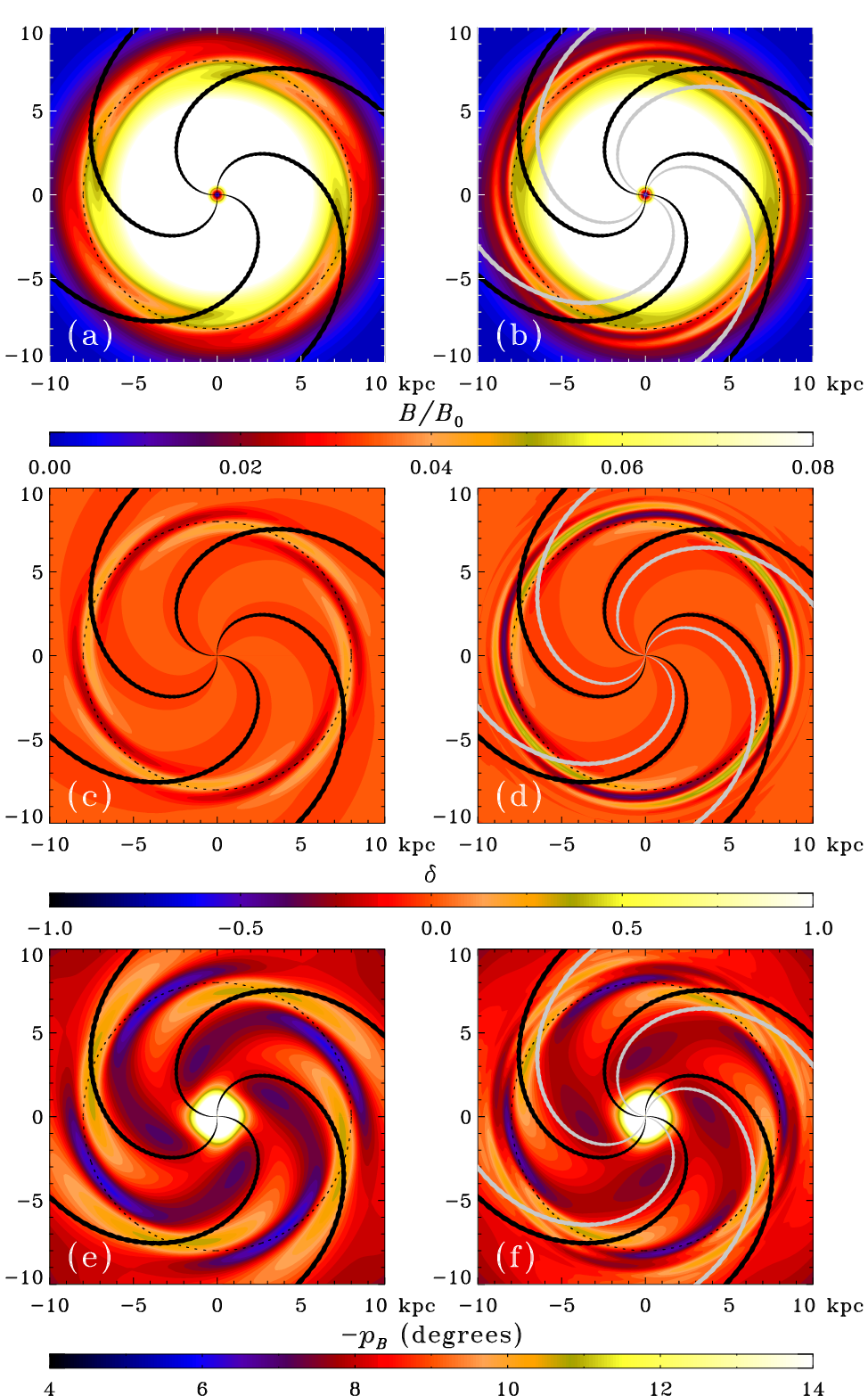}	
\caption{\small 
Same as Fig.~\ref{fig:kappa8_vec_dynq} but now for Model~L ($k=-20R^{-1}$ and $n=4$ instead of $2$).
\label{fig:narm4_vec_dynq}}
\end{center}
\end{figure}

We illustrate in Fig.~\ref{fig:sf_kin} 
various aspects of the solution in the kinematic stage at $3t\f$ after the simulation has begun (or $t=-5.8\Gyr$ in the plots).
Some aspects of the solution after the dynamo has saturated are given in Fig.~\ref{fig:sf_dynq}.
In Figs.~\ref{fig:sf_kin}a and \ref{fig:sf_dynq}a, $\mbp$ (solid), $-\mbr$ (dashed) and $|\mbz|$ (dotted)
have been plotted for both $\tau\rightarrow0$ (red) and $\tau=l/u$ (black) cases, 
at $\phi=\phi\corot$, one of the two azimuthal angles where the $\alpha$-spiral crosses the corotation circle.
As expected, the field is strongest inside the corotation radius, 
where the magnitude of the dynamo number is largest.
Importantly, however, there is an excess of magnetic energy near the corotation radius,
as compared to what is obtained in an axisymmetric disc,  
mainly due to the presence of strong additional $m=0$ and $m=2$ components near $r=r\corot$.
This excess is present at early enough times in the kinematic stage (like we have shown in Fig.~\ref{fig:sf_kin}a),
disappears at later times, but reappears on saturation.

The presence of the non-axisymmetric component near corotation can be more clearly seen in Fig.~\ref{fig:sf_kin}b.
There we plot the ratio of the non-axisymmetric part of $B_\phi$ to the axisymmetric part,
\[
\delta=\frac{B_\phi-B_\phi^{(0)}}{B_\phi^{(0)}}.
\]
at the azimuth $\phi=\phi\corot$.
For both vanishing and finite $\tau$, the $\delta$ is important within an annulus of width $\approx4\kpc$, centred near the corotation circle.
Note that in both cases, the radial phase varies much more rapidly than that of $\alpha$ (plotted as a thin dash-dotted line in the figure).
This indicates that the magnetic arms are much more tightly wound than the material arms.
The effect of a finite $\tau$ is to strengthen the non-axisymmetry, and also to slightly increase the variation of the radial phase.
In addition, the $\delta$ extends out to somewhat larger radius in the $\tau=l/u$ case than it does for $\tau\rightarrow0$.
This is illustrated by the second maximum with $\delta>0$ at $r\simeq10\,{\rm kpc}$  that occurs when $\tau=l/u$,
and is caused by the tail of the spiral magnetic field wrapping around an extra half-circle from where it crosses the corotation circle.
Also shown in Figs.~\ref{fig:sf_kin}b and \ref{fig:sf_dynq}b as dashed lines is the envelope of the
solid lines as $\phi$ changes.\footnote{The envelope is not perfectly symmetrical about $\delta=0$ 
because of interference between the $m=2$ mode and the much smaller $m=4$ mode.}

The shift between the azimuthal positions where $\mbp$ (or $\mbr$) is maximum and where $\alpha$ is maximum, 
denoted $\Delta_\phi$ (or $\Delta_r$), is shown in Figs.~\ref{fig:sf_kin}c and \ref{fig:sf_dynq}c.\footnote{Phase 
shifts quoted have an uncertainty of $\pm3^\circ$ due to the finite numerical mesh.}
Positive values of these quantities imply that the magnetic arms lead the material arms,
while negative values imply that they trail them.
The position $r\ma$
of the global maximum of $\delta$ (maximum of the dashed envelope in Figs.~\ref{fig:sf_kin}b and \ref{fig:sf_dynq}b), 
which is the radius at which the non-axisymmetric component of the field is most important relative to the axisymmetric component,
is indicated with a vertical dotted line (red for $\tau\rightarrow0$ and black for $\tau=l/u$): 
$r\ma=8.3\kpc$ (kinematic) and $8.0\kpc$ (saturated) for $\tau\rightarrow0$,
and $8.7\kpc$ (kinematic) and $8.6\kpc$ (saturated) for $\tau=l/u$.
One can see that $r\ma$ is nearer $r\corot$ for $\tau\rightarrow0$ than for $\tau=l/u$.

In the $\tau\rightarrow0$ case, studied by \citetalias{ms91} and \citetalias{sm93},
it was found that $\Delta_\phi=\Delta_r=0$ at $r=r\corot$ and $r\ma=r\corot$.
In the vanishing $\tau$ case in our model, $r\ma$ is slightly larger than $r\corot$ in the kinematic regime, 
and $\Delta_\phi$ and $\Delta_r$ are not precisely zero at $r\ma$ nor at $r\corot$. 
These differences arise due to interference from the dominant axisymmetric eigenmode, 
which is concentrated at smaller radius near where the dynamo number peaks, in our model.
In any case, for both $\tau\rightarrow0$ and $\tau=l/u$, the magnetic and $\alpha$-spiral arms intersect in the vicinity of corotation,
and the magnetic arms are also strongest in the vicinity of corotation, as in the earlier models.

What is more noteworthy, however, is the phase shift in the magnetic arms, at $r\ma$, where they are strongest, resulting from a finite $\tau$.
We find, initially, $\Delta_\phi(r\ma)=\Delta_r(r\ma)=-3^\circ$ for $\tau\rightarrow0$ and $-30^\circ$ for $\tau=l/u$, 
for an overall phase difference of $-27^\circ$.
On saturation, $\Delta_\phi(r\ma)=\Delta_r(r\ma)=+3^\circ$ for $\tau\rightarrow0$ and $-27^\circ$ for $\tau=l/u$,
for a phase difference of $-30^\circ$.
This is comparable in magnitude to the value $\Omega\tau=18^\circ$ that results from an order of magnitude estimate of the phase shift.
Non-locality in time produces a delay in the $\alpha$ effect of order $\tau$,
so that the $\alpha$-spiral has rotated through an angle $\approx\Omega\tau$ before the dynamo has had the chance to respond to it.
We recall that a phase shift of $\pm90^\circ$ would correspond to peaks of the magnetic arms located mid-way between the spiral arms of $\alpha$. 

The magnetic pitch angle
\[
p_B=\arctan\frac{\mbr}{\mbp}
\]
is plotted (by magnitude) in Figs.~\ref{fig:sf_kin}d and \ref{fig:sf_dynq}d, with its azimuthal mean shown as a dashed line.
The negative value of $p_B$ indicates that magnetic lines have the form of a trailing spiral. 
The azimuthally averaged value of $|p_B|$
is only slightly perturbed from the corresponding curve obtained in the axisymmetric case (i.e. for Model~B, not shown).
In the kinematic regime, the azimuthal average decreases with radius such that $\tan p_B \propto 1/h$,
as expected \citep{rss88,sh05}.
However, this decreasing trend is no longer present in the nonlinear regime, 
and the azimuthal average of $p_B$ (dashed curve) is almost constant over a large range of radius.

We have also checked the validity of neglecting the terms $\del(\alpha\mbz)/(r\del\phi)$ and $\del(\alpha\mbz)/\del r$ 
in Eqs.~\eqref{tau_r} and \eqref{tau_phi}, respectively.
Although neglecting the former is always justified, 
we find that the latter is comparable to the term $\del(\alpha\mbr)/\del z\simeq-2\alpha\mbr/(\pi h)$ in magnitude 
for some values of $\phi$ near $r=r\corot$, 
where radial variation of the solution is especially strong.
But this term itself, which appears in the equation for $\mbp$, 
is subdominant to the shear term $G\mbr$.
Therefore we repeated the run with the $\alpha\omega$ approximation (neglecting the $\alpha^2$ effect); 
we find the resulting differences in the solutions to be small and inconsequential.

Figure~\ref{fig:vec_dynq} shows a 2D representation of the magnetic field strength (upper panels), the quantity $\delta$ (middle panels), 
and pitch angle (lower panels) obtained in Model~E, 
with the case of $\tau\rightarrow0$ on the left, $\tau=l/u$ in the middle column, and $\tau=2l/u$ on the right.
Magnetic arms are clearly visible near the corotation circle in Fig.~\ref{fig:vec_dynq}a--f.
As they are more tightly wound than the $\alpha$-arms, they cut across them, as noted above.
From Fig.~\ref{fig:vec_dynq}, we see that the magnetic arms are more pronounced (larger $\mean{B}$ at their centres), 
are more sharply defined (faster variation of the radial phase),
and extend for a longer azimuthal angle outside of the corotation circle 
for the finite $\tau$ case compared to the $\tau\rightarrow0$ case.
Also, the maxima in magnetic field strength are displaced downstream from those of $\alpha$
by an angle of order $\Omega\tau$.
Moreover, in the $\tau=l/u$ case, a more significant part of the magnetic arm clearly lies in between the $\alpha$-spiral arms.
These features can be seen more clearly for $\tau=2l/u$ in the right hand panels of Fig.~\ref{fig:vec_dynq}.
It is worth noting that we find an enhancement in the amplitude of $\alpha\magn$ around where the non-axisymmetric mode concentrates, 
as required to saturate the dynamo.
This is visible in Fig.~\ref{fig:alpha_m}a and b, which show the profiles of $\alpha$ and $\alpha\magn$, respectively.

There is a hint of wave-like behaviour in the solutions with finite $\tau$, 
which is most evident in Fig.~\ref{fig:vec_dynq}f, 
where ripple-like features are visible just outside of the corotation circle.
But in general, $\tau$ is not large enough in our model for such wave-like behaviour to dominate.

The magnitude of the pitch angle is displayed in the lower panels of the figures, and clearly also 
has a spiral pattern.
The pitch angle, as opposed to the field magnitude, does not vary strongly with radius.
Because of this, the spiral morphology of the field is more clearly visible in the pitch angle.
However, an accuracy of a few degrees would be required to measure the differences.

Another quantity that can be obtained is the pitch angle of the magnetic ridges,
i.e. of the magnetic arms themselves as opposed to the magnetic field that constitutes them.
This can then be compared with the pitch angle of the $\alpha$-spiral arms.
We find this pitch angle to be quite small as compared to that of the $\alpha$-spiral.
One must keep in mind, however, that the pitch angle of the magnetic ridges is sensitive to the shear near corotation.

In Model~P, the run described in this section was performed with the resolution doubled for the cases $\tau=l/u$ and $2l/u$, 
and the results were found to be consistent with those of the standard resolution runs.

In Model~O, the advective flux of $\alpha\magn$ was replaced with the diffusive flux, 
and the results in the steady state are illustrated in Fig.~\ref{fig:vec_dynq_diffusive}.
The magnetic arms are almost identical (middle row), but the overall field strength is about twice as large (top row; as discussed in Sect.~\ref{sec:flux}), 
and magnetic field pitch angles are slightly smaller (bottom row).
The strength of the saturated field just outside of the corotation radius falls off somewhat more slowly with radius than for Model~E, 
which gives the {\it appearance} of wider magnetic arms in the top row of Fig.~\ref{fig:vec_dynq_diffusive}.
The enhancement of $\alpha\magn$ near corotation, mentioned above, 
is still present, though the feature is smoothened somewhat by the diffusive flux.
We emphasize that the qualitative features of the solutions obtained in both cases (Models~E and O) are very similar.
\subsection{Exploring alternative models and the parameter space}
\begin{figure}
\begin{center}
\includegraphics[width=84mm]{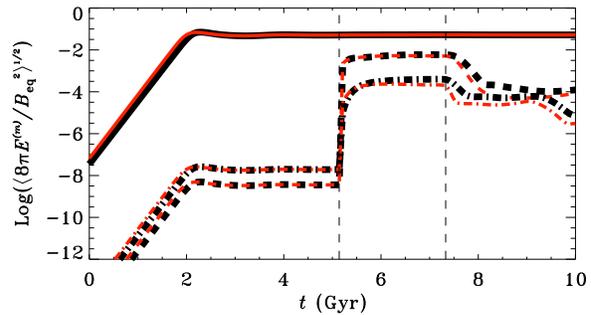}	
\caption{ 
Same as Fig.~\ref{fig:Gam}, but for Model~M (transient rigidly rotating spiral).
The times at which the $\alpha$-spiral was turned on and off are indicated by vertical dashed 
lines. 
\label{fig:transient_Gam}}
\end{center}
\end{figure}
\begin{figure*}
\includegraphics[width=175mm]{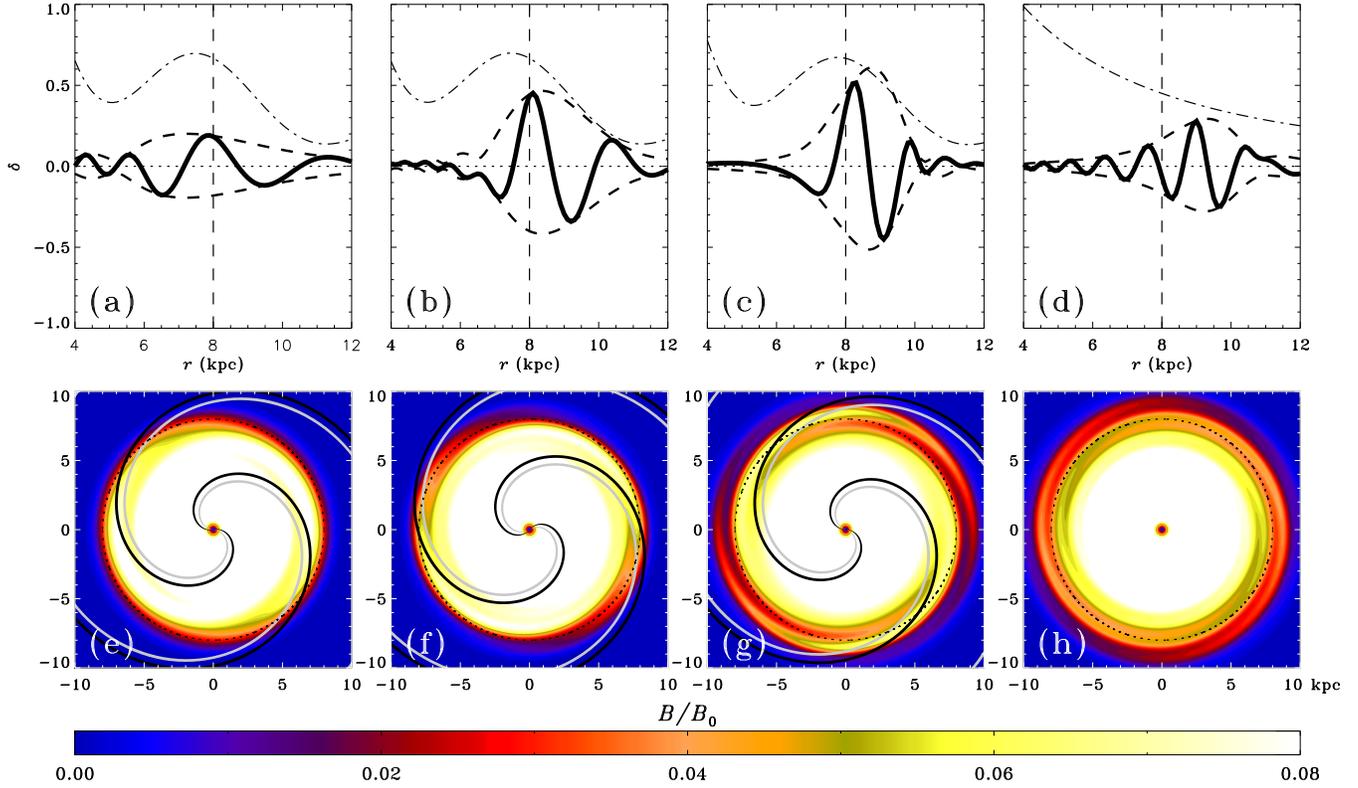}
\caption{ 
Evolution of the magnetic field under the action of a transient, rigidly rotating $\alpha$-spiral 
(Model~M) for $\tau=l/u$.
Each column shows the ratio of non-axisymmetric to axisymmetric part of $\mbp$, as in Fig.~\ref{fig:sf_kin}b (top) and magnitude of the field as in Fig.~\ref{fig:vec_dynq}b (bottom)
at the following times after the emergence of the $\alpha$-spiral:
{\bf (a, e)}~$0.22\,{\rm Gyr}$,
{\bf (b, f)}~$0.44\,{\rm Gyr}$, 
{\bf (c, g)}~$2.2\,{\rm Gyr}$, followed by
{\bf (d, h)}~$0.37\,{\rm Gyr}$ after the $\alpha$-spiral is turned off.
\label{fig:dynq_361_vec}}
\end{figure*}
Modulation of the $\alpha$ effect is just one of the mechanisms through which the spiral pattern can affect the dynamo action. 
Model~F has the vertical advection velocity enhanced along a spiral (with the $\alpha$-spiral 
turned off), to model stronger galactic outflows from the spiral arm regions, where star formation is enhanced. 
We use $R_U=0.90$ at $r=R/2=10\kpc$ and $\epsilon_U=0.5$ so that the outflow speed within the arms, 
corresponding to $R_U=1.35$, is too large for the optimal field growth near corotation, 
while the inter-arm value of $R_U=0.45$ is close to being optimal.
This mechanism for producing magnetic arms located between the material spiral arms was first 
suggested by \citet{sss07}.
We find that this mechanism can indeed lead to a non-axisymmetric field which peaks in the inter-arm regions. 
For the region of parameter space explored, the effect is, however, 
too weak to be of much consequence, though it might still be important for different parameter values.

The characteristic strength of magnetic field at which nonlinear dynamo effects become pronounced,
$B\eq$, can also be affected by the spiral pattern, e.g., through the variation in the turbulent
energy density. We considered a model with $B\eq$ enhanced along a spiral in the same way as above 
for $\alpha\kin$ and $\muz$. Since $B\eq=(4\pi\rho)^{1/2}u$, and we keep $u$ constant, this is 
tantamount to
modulating $\rho$. Although an enslaved spiral does result, we find the effect to be 
too weak to be of much consequence, at least for the parameter space explored.
Other possibilities have been explored in the literature, e.g. the modulation of the turbulent 
magnetic diffusivity $\eta\turb$, but a more extensive study of these effects is left for a future work.

We have also run Models~G--L to test the results under reasonable variations of parameters. 
We have varied the strength of the seed field, the mean vertical velocity $\muz$, 
the strength and  the pitch angle of the spiral pattern via $\epsilon_\alpha$ and $k$,
respectively, and the number of material spiral arms $n$.
In Model~G, we use a seed field that is random Gaussian noise.
Model~G differs from Model~C in that the underlying disc in non-axisymmetric. 
We find reversals to occur in more or less the same locations {as in that model}.
However, 
a strong quadrisymmetric magnetic field component is now present, superimposed on an axisymmetric pattern with reversals.
The morphology of the magnetic spiral arms is significantly affected by the reversals,
as the horizontal magnetic field is necessarily zero wherever a reversal occurs.

The effect of vanishing helicity flux is 
considered in Model~H, which differs from Model~E only in that it has $\muz=0$.
This situation can occur in galaxies after the end of a burst of star formation,
when the galactic fountain or wind ceases and the mean-field dynamo action is choked
by the magnetic helicity conservation to leave magnetic field decaying,
assuming no other flux is important.
As expected, the mean field decays after a period of temporary growth.
An expanding annular magnetic structure (see Sect.~\ref{sec:flux}) is prominent, 
but develops strong deviations from axial symmetry as the ring approaches and then passes through the corotation circle. 
As this happens, the amplitude of $\alpha\magn$ becomes enhanced there, 
as can be seen in Fig.~\ref{fig:alpha_m}c-d, and then remains enhanced subsequently.
The magnetic ring fades in intensity and becomes more and more axisymmetric as it moves out.

In Model~I, $\epsilon_\alpha$ is increased from 0.5, as in Model~E, to $\epsilon_\alpha=1$,
so that now $\alpha\kin=0$ between the spiral arms.
As could be expected, this enhances the deviation of the mean magnetic field from axial symmetry, 
manifested in more pronounced magnetic arms and a larger azimuthal variation in the magnetic pitch angle.
The magnitude of the non-axisymmetric part of the magnetic field mode slightly exceeds the axisymmetric part near $r=r\ma$ when $\tau=l/u$.
This produces small regions of positive pitch angle just outside of the corotation,
where the magnetic lines locally have the shape of a leading spiral.
Putting $\tau=l/u$ with $\epsilon_\alpha=1$ also results in a somewhat larger phase shift compared to the $\tau=l/u$, 
$\epsilon_\alpha=0.5$ case in Model~E.
We obtain $\Delta_r(r\ma)=\Delta_\phi(r\ma)=-36^\circ$ for $\tau=l/u$,
whereas for Model~E we had $-27^\circ$.
For $\tau\rightarrow0$, the phase shifts are both equal to $+3^\circ$, 
the same as in Model~E so the overall phase difference is $-39^\circ$, larger than for Model~E.
The trailing part of the magnetic spiral is significantly enhanced when $\tau$ is finite.

In Models~J and K we make the $\alpha$-spiral less tightly wound by reducing the magnitude of $k$.
It is evident from Fig.~\ref{fig:kappa8_vec_dynq} that
changing the value of $k$ from $-20R^{-1}$ 
[$p_\alp(r\corot)=-14^\circ$, Model~E] to $-8R^{-1}$ [$p_\alp(r\corot)=-32^\circ$, Model~J]
does not have any significant qualitative effect on the magnetic field.
Even replacing the material spiral with a bar ($k=0$, Model~K) 
does not lead to a drastic change in the saturated magnetic field,
so that bars also lead to spiral magnetic fields.
This happens because of the differential rotation of the gas, 
which shears out the enhancement of the field due to the bar.
An important effect of a more open spiral pattern, visible in Fig.~\ref{fig:kappa8_vec_dynq}a--b,
is that magnetic spiral arms are now mostly in between the material arms, 
although still confined to an annular region around corotation.
The values of $\Delta_r(r\ma)$ and $\Delta_\phi(r\ma)$ are both $-33^\circ$ for $\tau=l/u$,
and $+3^\circ$ for $\tau\rightarrow0$, for an overall phase difference of $-36^\circ$. 
The pattern of variation of the magnetic pitch angle is strongly modified from the case of the more tightly wound $\alpha$-spiral of Model~E, 
though its range is almost the same.

We tried various values of $n$, both odd and even, and found, unsurprisingly, that the number of magnetic arms is 
equal to the number of material arms.
The results of Model~L ($n=4$) are plotted in Fig.~\ref{fig:narm4_vec_dynq}.
The magnetic arms are largely located in between the material arms, especially for $\tau=l/u$.
They are also much stronger and more well-defined in the finite $\tau$ case.

The possibility of `mode-switching' in the nonlinear regime has been pointed out by \citet{hubbardetal11}.
We do not find such solutions in the present framework, 
but it would be worthwhile to revisit this issue in the context of three-dimensional models which include the galactic halo.

\section{Magnetic response to transient spiral patterns}\label{TSP}\label{sec:transient}
\subsection{A rigidly rotating spiral pattern}
As galactic spiral patterns may be transient in nature, we now explore, in Model~M, 
how the magnetic field responds to sudden changes in the spiral forcing.
The runs are identical to Model~B (axisymmetric disc) up until the $\alpha$-spiral is turned on, 
at the time $t\on=5.1\Gyr$, when the dynamo is already in its nonlinear phase. 
Subsequently, the parameters of the model are identical to those of Model~E (standard spiral 
forcing) up until the time $t\off=7.3\Gyr$, when the spiral modulation of $\alpha$ 
is turned off.

The evolution of the magnetic field strength in each Fourier mode, 
averaged over the area of the disc, is shown in Fig.~\ref{fig:transient_Gam}. 
The $m=2$ mode, which initially is present solely because of numerical noise, 
responds rapidly to the onset of the $\alpha$-spiral, 
and then behaves almost exactly as in Model~E (compare with Fig.~\ref{fig:Gam}). 
In fact, for $\tau=l/u$, magnetic energy in the $m=2$ part of the mean field
slightly exceeds that in Model~E before being reduced to the latter.
The timescale of the adjustment of the magnetic field,
i.e., the time from the onset of the $\alpha$-spiral, taken for the $m=2$ part in Model~M to 
grow up to that in Model~E, is as short as $\simeq0.2\Gyr$. 

Figure~\ref{fig:dynq_361_vec} shows characteristic magnetic configurations for 
$\tau=l/u$ (the $\tau\rightarrow0$ case is qualitatively similar so only $\tau=l/u$ is shown). 
After a short period of $0.22\Gyr$ after the onset of the spiral forcing, 
the deviation from axial symmetry in the magnetic field is strongest somewhat inside of the corotation circle, 
(see Fig.~\ref{fig:dynq_361_vec}a).
This is explained by the dynamo responding more rapidly to sudden changes in the disc at locations where the dynamo number is larger.
At $t=t\on+0.44\Gyr$,
the non-axisymmetric component of the field has expanded outward in radius, 
and has become stronger than it was (as compared to the axisymmetric component; panels b and f).
This is to be expected since, according to results already discussed, deviations from axisymmetry are most important near the corotation circle.
Next, at $t=t\on+2.2\Gyr=t\off$ (Fig.~\ref{fig:dynq_361_vec}c and g), 
the magnetic field has reached nearly the same equilibrium state as in Model~E (compare with Figs.~\ref{fig:sf_dynq}b and \ref{fig:vec_dynq}b).
After the $\alpha$-spiral has been switched-off, 
the magnetic arms remain for a rather long time compared with the time it had taken for them to arise.
This can be seen in Figs.~\ref{fig:dynq_361_vec}d and h, where we show the field at $t=t\off+0.37\Gyr$.
The energy in the $m=2$ part declines to a quarter of its value at $t\off$ within $0.2\Gyr$ and $0.3\Gyr$, 
respectively, for $\tau\rightarrow0$ and $\tau=l/u$.
Therefore, `ghost' magnetic spiral arms remain after the demise of the material spiral arms, 
and survive longer when $\tau$ is finite.
We find an extra delay of about $0.1\Gyr$ due to the finite $\tau$,
larger than the delay $\tau\simeq0.01\Gyr$ that might naively be expected.
Similar results were seen in \citet{otmianowska-mazuretal02} using a quite different model.
\subsection{A winding-up spiral pattern} 
\begin{figure}
\begin{center}
\includegraphics[width=84mm]{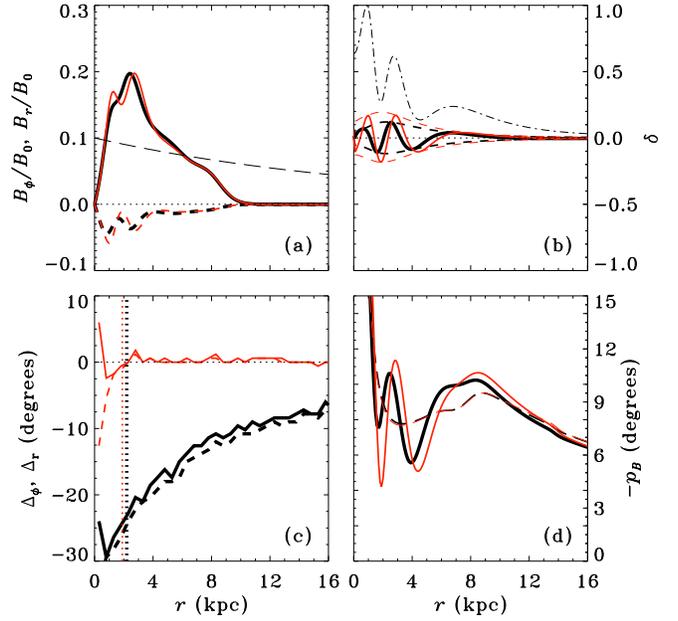}
\caption{Similar to Fig.~\ref{fig:sf_dynq}, 
but now for Model~N (the $\alpha\kin$-spiral winding up with the gas).
The figure shows the field at $0.1t\f=73\,{\rm Myr}$ 
after the sudden onset of an $\alpha\kin$ bar 
when the dynamo action has already saturated.
\label{fig:sf_dynq_wind1}}
\end{center}
\end{figure}
\begin{figure}
\begin{center}
\includegraphics[width=84mm]{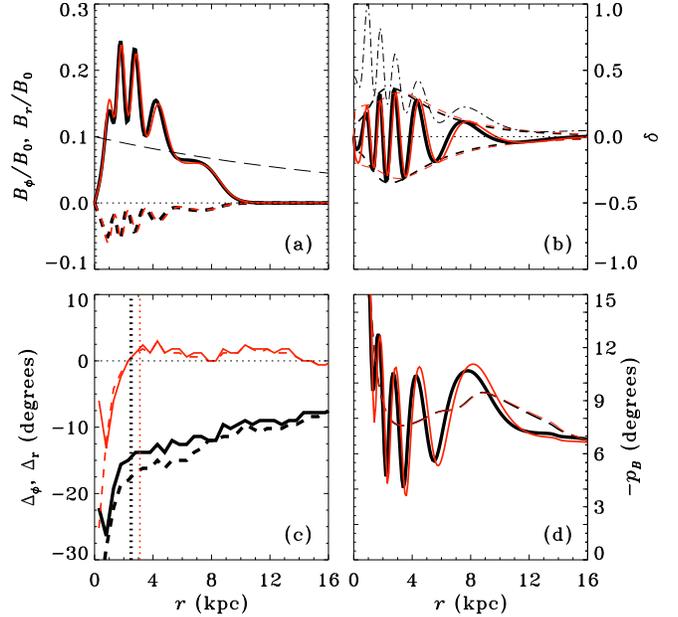}
\caption{Same as Fig.~\ref{fig:sf_dynq_wind1}, but at $0.2t\f=146\,{\rm Myr}$ 
after the onset of the $\alpha\kin$ bar (which subsequently winds up into a spiral). 
\label{fig:sf_dynq_wind2}}
\end{center}
\end{figure}
\begin{figure*}
\begin{center}
\includegraphics[width=132mm]{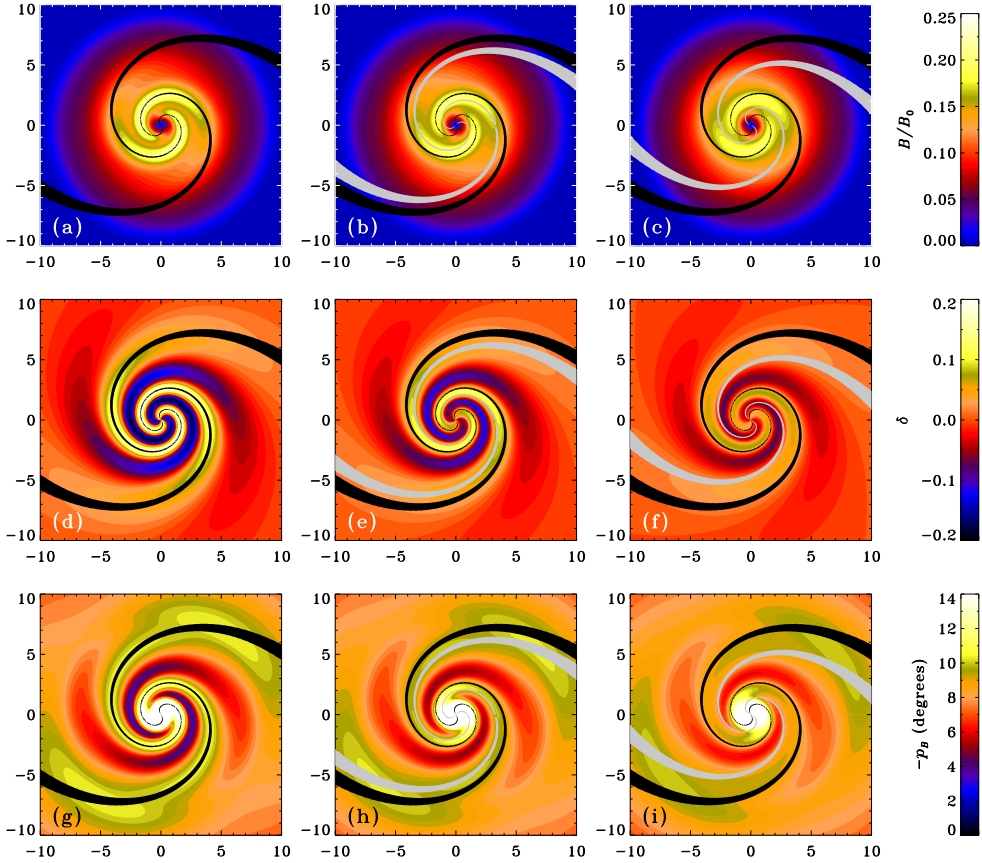}	
\caption{Winding-up spiral of Model~N. 
The figure is similar to Fig.~\ref{fig:vec_dynq} but now grey spirals represent $\alpha\kin$ maxima shifted by $-\omega\tau$.
All panels are snapshots at $0.1t\f=73\,{\rm Myr}$ after 
the  sudden onset of the $\alpha\kin$ `bar'.
\label{fig:vec_dynq_wind1}}
\end{center}
\end{figure*}
The models described so far had a rigidly rotating spiral with a constant pattern speed.
As discussed in the introduction, models with a spiral winding up with the differential rotation
may be more suitable for some galaxies -- 
in any case rigidly rotating steady spirals and winding-up transient spirals plausibly represent two extreme cases.

We now consider the case of an $\alpha$-spiral 
that winds up with the gas.
In this model every radius effectively becomes a corotation radius, 
so based on our results for a rigidly rotating spiral, we would expect the non-axisymmetric part of the magnetic field to be more extended radially, 
as seen in some observations.
In Model~N, the field evolves as in Model~B (axisymmetric disc) until the time $t\on$.
After $t\on$ we switch on the $\alpha\kin$ modulation,
but now let the spiral wind up starting from the `bar' stage ($k=0$) at $t=t\on=6.6\Gyr$, 
when the $m=0$ field is already saturated.
We find that strong deviations from axial symmetry in the magnetic field develop during the first $\sim150\Myr$ after $t\on$.
In Figs.~\ref{fig:sf_dynq_wind1} and \ref{fig:sf_dynq_wind2} we show the properties of the resulting field at two times, 
$t=t\on+0.1t\f=t\on+73\Myr$ and $t\on+0.2t\f=t\on+146\Myr$, respectively.

As can be seen by comparing panels a and b of Fig.~\ref{fig:sf_dynq_wind1} ($t=t\on+73\Myr$) 
and also by comparing panels a and b of Fig.~\ref{fig:sf_dynq_wind2} ($t=t\on+146\Myr$),
the non-axisymmetric and axisymmetric parts of the magnetic field are the strongest 
at about the same radii, so that the magnetic field is essentially non-axisymmetric.
This can also be seen from the 2D plot of Fig.~\ref{fig:vec_dynq_wind1} ($t=t\on+73\Myr$).
In addition, the magnetic spiral arms are extended throughout the disc over a much larger range of radii 
than with a rigidly rotating spiral pattern
(where the non-axisymmetric field concentrates around the corotation circle).

Figures~\ref{fig:sf_dynq_wind1}b and \ref{fig:sf_dynq_wind2}b show that $\delta$
(solid lines) and $\alpha\kin$ (dash-dotted line) are well-correlated, 
demonstrating that the dynamo responds rather quickly to the spiral forcing.
From Figs.~\ref{fig:sf_dynq_wind1}c and \ref{fig:sf_dynq_wind2}c, we see that for $\tau\rightarrow0$
the phase difference between the magnetic and material spiral arms 
is negligible at radii where the magnetic spiral arms are strong.
However, for finite $\tau=l/u$, the magnetic arms lag the material ones by a large angle, 
$\Delta_r\simeq\Delta_\phi\simeq-15^\circ$ to $-25^\circ$ over a range of radius of order $10\kpc$.
The values vary somewhat with radius and with time.
This phase shift is strikingly apparent when comparing panels (a) and (d) of Fig~\ref{fig:vec_dynq_wind1} for $\tau\rightarrow0$ with, 
respectively, panels (b) and (e) for $\tau=l/u$.
Only in panels (b) and (e) do the yellow-orange regions of high field strength lag the black centres of the $\alpha$-spiral arms by a significant amount 
but overlap with the grey spiral phase-shifted by the angle $-\omega\tau$.

As might be expected, the phase shift is larger for $\tau=2l/u$ ($-20^\circ$ to $-30^\circ$),
as can be seen from the third column of Fig.~\ref{fig:vec_dynq_wind1} (but not as large as $2\omega l/u$).
However, as $\tau$ is increased, the non-axisymmetric part of the magnetic field tends to weaken, as
the spreading of the response of the dynamo in time due to the finite dynamo relaxation time is translated into an 
effective blurring of the $\alpha$-spiral in space 
(the effect is most evident when comparing panels d--f of Fig.~\ref{fig:vec_dynq_wind1}).
When the advective flux is replaced by a diffusive flux with $\kappa=0.3\eta\turb$, we find almost identical results, 
save for the fact that the strength of the field is larger by about 75\%.

This model, which incorporates a spiral pattern that seems to wind up with the differentially rotating gas,
along with a dynamo relaxation time $\tau\gtrsim l/u$,
seems to be the most successful, among the models explored in the present work, 
at explaining the large phase shifts between magnetic and material arms, roughly constant over a large radial range, observed in some galaxies,
such as NGC~6946.
\section{Discussion and conclusions}
\label{sec:conclusion}
We have presented a set of models of the mean-field galactic dynamo action that include:
(i)~dynamo forcing by stationary and transient spiral structure,
(ii)~time delay in the response of the mean electromotive force to changes in the mean magnetic 
field and small-scale turbulence (finite relaxation time $\tau$), and
(iii)~dynamo nonlinearity based on magnetic helicity balance. 

Whilst some aspects of (i) have been explored earlier in simpler galactic dynamo models, 
the latter two features present significant generalizations of the galactic dynamo theory which, 
we believe, make it more realistic and physically  appealing. 
In particular, (ii) affects the mathematical nature of the mean-field dynamo equation, 
which now has the form of the telegraph equation admitting diffusive wave-like solutions. 
Solutions presented here do not have any pronounced wave-like properties, 
but this does not preclude that such solutions of the generalized dynamo equation do exist. 
Feature (iii) incorporates a physically justifiable form of nonlinearity into galactic dynamo equations 
(in the form of advective or diffusive helicity fluxes), 
thus making our solutions significantly more realistic than those explored earlier.

Within this framework, we explore the behaviour of the mean magnetic field in both axisymmetric and non-axisymmetric galactic discs.
A novel aspect of this paper is the introduction of a finite relaxation time into the galactic 
mean-field theory and a detailed exploration of its effects, particularly on the non-axisymmetric dynamo modes.
We discuss how the imprints of the 
galactic spiral pattern on the morphology of the mean magnetic field are affected by the finite 
relaxation time of the dynamo. 
Unlike many earlier studies of galactic dynamos, we approach these 
problems from the evolutionary viewpoint and consider the magnetic response to both steady and 
transient spiral patterns, and briefly discuss the diverse ways in which galactic spiral arms can 
affect the dynamo action.

We have confirmed, in the new framework, earlier results on the evolution of axisymmetric magnetic fields in axisymmetric discs, 
including the spreading of magnetic fronts along radius and the occurrence of reversals.
As expected, the magnetic field is able to spread radially to encompass the whole disc within the galactic lifetime,
as long as the magnetic helicity flux is sufficient.
What is new is that in the absence of any helicity flux, 
contracting and expanding rings of magnetic field, emanating from the region where the magnitude of the dynamo number is largest,
can propagate for several Gyr in the nonlinear regime.
This is due to the variation along radius of the time taken for the magnetic field to reach saturation.

The lifetimes of reversals in the galactic disc have been known to be sensitive to the details of the galactic rotation curve, 
the geometric shape of the gas layer, details of the turbulent energy distribution in the galaxy, etc. 
We have added another important parameter to this list, the dynamo relaxation time. 
We also find that such reversals can have non-trivial effects on the morphology 
of non-axisymmetric magnetic structures in a non-axisymmetric disc.
It would be interesting to carry out a more detailed study of magnetic fronts and the possibility of long-lived reversals 
in the context of the above framework.

We have also shown that it is difficult to maintain non-axisymmetric magnetic structures in an axisymmetric disc,
which leaves one little choice but to try to explain such structures as arising from a non-axisymmetric disc.
Most of our effort was then directed to the response of the mean magnetic field to the galactic spiral 
pattern. The focus in this part of our work is the nature of the so-called magnetic arms, 
spiral-shaped regions of enhanced regular magnetic field (traced by polarised radio emission and Faraday rotation) 
that are located in between the gaseous and stellar spiral arms 
in some galaxies (e.g., NGC~6946 and IC~342) but can overlap with or cross the material arms in other galaxies (e.g., M51). 
We model the effect of the spiral pattern on the dynamo by assuming that the 
$\alpha$ effect or the galactic/fountain outflow or the equipartition field are enhanced within the spiral arms. 
A non-axisymmetric $\alpha$ effect, in particular, does lead to strong magnetic arms.

When the dynamo relaxation time $\tau\rightarrow0$, the magnetic and material arms almost coincide at corotation,
as expected from earlier work \citepalias{ms91,sm93,mo96}.
However, a finite relaxation time causes a significant
azimuthal lag between the magnetic and material arms at the radius where the magnetic arms are strongest (near corotation).
In this respect, our model can explain a wider range of observations than earlier models. 
However, we concur with the earlier authors in that the non-axisymmetric parts of the mean magnetic field driven by the spiral 
pattern are mainly localised around the corotation radius, extending about $4\kpc$ in radius. 
The radial extent is, of course, model-dependent and can be larger in regions with weaker rotational velocity shear. 

The corotation radius is also approximately where each magnetic 
arm crosses the corresponding material arm, going from leading (with respect to the direction of 
the galactic rotation) the material arm inside the corotation circle to trailing it outside the 
corotation.
The primary effect of the finite relaxation time, in the case of a steady, rigidly rotating spiral
pattern, is to suppress the leading part of the magnetic spiral arms and to enhance, 
and extend azimuthally, the trailing part. 
Thus, this effect produces a prominent  `tail' of large-scale magnetic field in between the material arms.
This is even more true when the number of arms is increased from two, to say, four.

The magnetic arms of the spiral galaxy NGC~6946 are `interlaced with' (in between) the material arms \citep{behoernes96,be07}, 
and have been called `phase-shifted images' of the preceding (in the sense of the rotation) material arms \citep{fricketal00}.
Although the implied phase shifts are more constant with radius than in our model, 
we have obtained a shift of approximately the same magnitude ($30^\circ$--$40^\circ$) and in the right direction.
More generally, the shift is of order $\Omega\tau$, where $\Omega$ is the pattern angular velocity of the material spiral.
Larger values of $\tau$ thus lead to larger phase shifts.
This is caused by the delay in the $\alpha$ effect of order $\tau$,
so that by the time the dynamo has had the chance to respond to the enhancement in $\alpha$ along the material arm,
the arm has already rotated by an angle $\approx\Omega\tau$.

Spiral density waves may be transient and we consider how quickly the regular magnetic field can 
respond to changes in the galactic spiral structure.
We find that the response time is small.
On the other hand, 
we find that magnetic spiral arms survive for several hundred Myr 
following the destruction of the material spiral arms in the dynamo nonlinear regime.
Moreover, a finite dynamo relaxation time is found to significantly prolong the life of such lingering magnetic arms. 
This opens the intriguing possibility of `ghost' magnetic arms, 
which were produced by material arms that have since disappeared.

Despite the wide range of models considered, our success in reproducing magnetic arms interlaced
with the material arms as perfectly as it is believed to happen in NGC~6946 is admittedly limited
in models assuming forcing of the dynamo by a steady, rigidly rotating spiral. 
In addition, this type of spiral forcing leads to magnetic arms concentrated over a smaller range in radius than is observed in many galaxies.
Another possibility, rendered more likely by several recent studies \citep[e.g.][]{dobbsetal10, sellwood11, quillenetal11},
is that the spiral patterns of many galaxies, 
rather than being rigidly rotating, as is usually assumed in galactic dynamo models, in fact wind up (at least to some extent).
This may happen if, for instance, there are interfering two and three-armed spirals rotating at different angular frequencies
(A. Quillen, private communication).
The two-arm grand design spiral pattern of the galaxy M51 (NGC~5194), 
thought to be caused by tidal forcing by its neighbour, NGC~5195, 
is also found to wind up in detailed N-body simulations that are able to accurately reproduce its spiral morphology \citep{dobbsetal10}. 

With these recent advances in spiral structure theory in mind, we also investigated the opposite extreme to rigidly rotating patterns: 
material arms that are wound up by the galactic differential rotation. 
The nonlinear dynamo responds very quickly to such forcing, 
and for vanishing dynamo relaxation time the mean magnetic field more or less traces the spiral arms 
over a large range in radius (as seen in many observations) and winds up with them.
On the other hand, for finite dynamo relaxation time
there is a large azimuthal lag of each magnetic spiral arm compared to the corresponding material arm over a large range in radius. 
Magnetic arms trail the material arms by $15^\circ$--$25^\circ$ (for $\tau=l/u$), varying somewhat with time and radius over the disc.
This shift is of order $\omega\tau$, where $\omega$ is the angular velocity of the gas, and we have shown that larger, 
but still plausible, values of $\tau$, lead to even larger phase shifts.
Increasing the number of spiral arms also causes the (equal number of) magnetic arms to be located closer to the centres of the inter-arm regions,
so that they may be described as interlaced with the material arms.
Therefore, allowing for the possibility of the spiral winding up can drastically improve agreement 
with observations of the regular magnetic fields in some galaxies, 
but with the trade-off that the magnetic arms (that in this model, either trace or interlace the material arms) 
are almost as short-lived as the spiral patterns that presumably generate them.

In summary, we have 
incorporated into galactic dynamo theory several well-studied physical effects not previously considered,
namely (i) non-locality in time and 
(ii) forcing by both spiral arms which steadily rotate and those which wind up due to differential rotation.
This allows several observed features of magnetic arms to be more naturally reproduced.
Particularly interesting are the models we have presented in which magnetic arms extend over a large range of radii
and either trace material arms over several kpc, or else are phase-shifted images of material arms, 
trailing them in the sense of the galactic rotation.
It would be interesting and important to test these ideas by applying them to specific galaxies, for which the rotation curve,
velocity dispersion, spiral structure, etc., can be constrained by existing data from observation and simulation.

\section*{Acknowledgements}
We are grateful to Sharanya Sur for sharing with us his preliminary results on incorporating MTA 
into the galactic mean-field dynamo equations and for many useful discussions.
LC wishes to thank Axel Brandenburg for initial help with the simulations and for generously sharing his many routines, 
as well as for useful discussions. 
We also thank Nishant Singh for reading an early draft of the manuscript and providing valuable suggestions. 
AS is grateful to IUCAA for financial support.
KS acknowledges partial support from NSF Grant PHY-0903797 while at the University of Rochester. 
KS thanks Eric Blackman at Rochester for warm hospitality and both him and Alice Quillen for interesting discussions.
We thank the referee for insightful comments and suggestions that helped to improve the paper.

\appendix
\section{A more general form of the basic equations}\label{EGF}
\label{sec:eqns_general}
\subsection{The telegraph equation}
Having defined
\[
\Fmf\equiv \Del\cro\Emf,
\]
and assuming that $\eta$, $\eta\turb$, $\tau$ and $c_\tau$ are constants,  
the cylindrical polar components of Eq.~\eqref{meaninduction} can be written as
\begin{equation}
\label{ind_r}
\begin{split}
\frac{\del\mbr}{\del t}=&\frac{1}{r}\frac{\del}{\del\phi}(\mur\mbp-\mup\mbr)-\frac{\del}{\del z}(\muz\mbr-\mur\mbz)+\Fmfr\\
&+\eta\left\{\frac{\del}{\del r}\left[\frac{1}{r}\frac{\del}{\del r}(r\mean{B}_r)\right]+\frac{1}{r^2}\frac{\del^2\mean{B}_r}{\del\phi^2}+\frac{\del^2\mean{B}_r}{\del z^2}-\frac{2}{r^2}\frac{\del\mean{B}_\phi}{\del\phi}\right\},\\
\end{split}
\end{equation}
\begin{equation}
\label{ind_phi}
\begin{split}
\frac{\del\mbp}{\del t}=&\frac{\del}{\del z}(\mup\mbz-\muz\mbp)-\frac{\del}{\del r}(\mur\mbp-\mup\mbr)+\Fmfphi\\
&+\eta\left\{\frac{\del}{\del r}\left[\frac{1}{r}\frac{\del}{\del r}(r\mean{B}_\phi)\right]+\frac{1}{r^2}\frac{\del^2\mean{B}_\phi}{\del\phi^2}+\frac{\del^2\mean{B}_\phi}{\del z^2}+\frac{2}{r^2}\frac{\del\mean{B}_r}{\del \phi}\right\},\\
\end{split}
\end{equation}
\begin{equation}
\label{ind_z}
\begin{split}
\frac{\del\mbz}{\del t}=&\frac{1}{r}\frac{\del}{\del r}\left[r(\muz\mbr-\mur\mbz)\right]-\frac{1}{r}\frac{\del}{\del\phi}(\mup\mbz-\muz\mbp)\\
&+\Fmfz+\eta\left[\frac{1}{r}\frac{\del}{\del r}\left(r\frac{\del\mean{B}_z}{\del r}\right)+\frac{1}{r^2}\frac{\del^2\mean{B}_z}{\del\phi^2}+\frac{\del^2\mean{B}_z}{\del z^2}\right].
\end{split}
\end{equation}
Taking the curl of both sides of Eq.~\eqref{minimaltau2}, we get
\begin{equation}
\label{tau_r}
\begin{split}
\tau\frac{\del\Fmfr}{\del t}=&c_\tau\left[\frac{1}{r}\frac{\del}{\del\phi}(\alp\mbz)-\frac{\del}{\del z}(\alp\mbp)\right]+c_\tau\eta\turb\left\{\frac{\del}{\del r}\left[\frac{1}{r}\frac{\del}{\del r}(r\mean{B}_r)\right]\right.\\
&\left.+\frac{1}{r^2}\frac{\del^2\mean{B}_r}{\del\phi^2}+\frac{\del^2\mean{B}_r}{\del z^2}-\frac{2}{r^2}\frac{\del\mean{B}_\phi}{\del \phi}\right\}-\Fmfr,
\end{split}
\end{equation}
\begin{equation}
\label{tau_phi}
\begin{split}
\tau\frac{\del\Fmfphi}{\del t}=&c_\tau\left[\frac{\del}{\del z}(\alp\mbr)-\frac{\del}{\del r}(\alp\mbz)\right]+c_\tau\eta\turb\left\{\frac{\del}{\del r}\left[\frac{1}{r}\frac{\del}{\del r}(r\mean{B}_\phi)\right]\right.\\
&\left.+\frac{1}{r^2}\frac{\del^2\mean{B}_\phi}{\del\phi^2}+\frac{\del^2\mean{B}_\phi}{\del z^2}+\frac{2}{r^2}\frac{\del\mean{B}_r}{\del \phi}\right\}-\Fmfphi,
\end{split}
\end{equation}
\begin{equation}
\label{tau_z}
\begin{split}
\tau\frac{\del\Fmfz}{\del t}=&c_\tau\left[\frac{1}{r}\frac{\del}{\del r}(r\alp\mbp)-\frac{1}{r}\frac{\del}{\del\phi}(\alp\mbr)\right]\\
&+c_\tau\eta\turb\left[\frac{1}{r}\frac{\del}{\del r}\left(r\frac{\del\mean{B}_z}{\del r}\right)+\frac{1}{r^2}\frac{\del^2\mean{B}_z}{\del\phi^2}+\frac{\del^2\mean{B}_z}{\del z^2}\right]-\Fmfz.
\end{split}
\end{equation}

\subsection{Dynamical quenching}
\label{sec:dynq}
To evaluate the term containing $\Emf$ in Eq.~\eqref{dalpha_mdt}, we need the evolution equations for its components. 
From Eq.~(\ref{minimaltau2}), we find
\begin{align}
\label{Emfr}
\tau\frac{\del\Emfr}{\del t}&=c_\tau\alpha\mbr-c_\tau\eta\turb\left(\frac{1}{r}\frac{\del\mbz}{\del\phi}-\frac{\del\mbp}{\del z}\right)-\Emfr,\\
\label{Emfphi}
\tau\frac{\del\Emfphi}{\del t}&=c_\tau\alpha\mbp-c_\tau\eta\turb\left(\frac{\del\mbr}{\del z}-\frac{\del\mbz}{\del r}\right)-\Emfphi,\\
\label{Emfz}
\tau\frac{\del\Emfz}{\del t}&=c_\tau\alpha\mbz-c_\tau\eta\turb\left[\frac{1}{r}\frac{\del}{\del r}(r\mbp)-\frac{1}{r}\frac{\del\mbr}{\del\phi}\right]-\Emfz.
\end{align}
The evolution equations for $\Emf$ are redundant given those for $\Fmf$, since the latter could be obtained by taking the curl of the former.
However, we find it convenient to solve separately for $\Emf$ and $\Fmf$ (see Sect.~\ref{sec:numerical_treatment}).

\section{The no-\lowercase{$z$} approximation}
\label{sec:noz}
\subsection{Vertical diffusion}
Under the no-$z$ approximation, the second derivatives with respect to $z$ can be 
approximated as $\del^2\mean{B}_i/\del z^2\simeq -\pi^2\mean{B}_i/4h^2$ (where $i=r,\phi$), which gives the correct sign of the  diffusion term. 
This approximation can be derived from the one-dimensional eigenfunctions 
obtained from the perturbation theory \citep{sss07,ss08}.

\subsection{The $\alpha$ effect}
When applying the no-$z$ approximation to the terms $\del(\alpha \mean{B}_i)/\del z$ in Eqs.~\eqref{tau_r} and \eqref{tau_phi}, 
one must be careful about the sign. 
The sign must be chosen so that the $\alpha$ effect can contribute to a positive dynamo growth rate.
This logic leads to the adoption of the approximations
\begin{equation}
\nonumber-\frac{\del}{\del z}(\alpha\mbp)\simeq-\frac{|\alpha|\mbp}{h}, \quad \frac{\del}{\del z}(\alpha\mbr)\simeq-\frac{|\alpha|\mbr}{h}.
\end{equation}
The second of these is only relevant when the $\alpha^2$ effect is taken into consideration.

Furthermore, \citet{phillips01} has shown that the no-$z$ approximation can be made more accurate with the additional numerical factor $2/\pi$,
at least for the $\alpha\omega$ dynamo.
As a matter of symmetry, we include the same numerical factor in front of both $\alpha$ terms, so that we may write
\begin{equation}
-\frac{\del}{\del z}(\alpha\mbp)\simeq-\frac{2|\alpha|\mbp}{\pi h}, \quad \frac{\del}{\del z}(\alpha\mbr)\simeq-\frac{2|\alpha|\mbr}{\pi h}.
\end{equation}

It may be asked whether extending the no-$z$ approximation to include the $\alpha$ term in the evolution equation for $\mbp$ (through Eq.~\ref{tau_phi}),
as we have done here for the first time, actually helps to improve the accuracy of the solution.
To answer this, we compared the kinematic solutions obtained using simple two-dimensional (in $r-z$) $\alpha\omega$ 
and $\alpha^2\omega$ galactic dynamo model 
with those obtained from the corresponding one-dimensional (in $r$, with no-$z$) models.
Interestingly, we found significantly better agreement between the 1D and 2D solutions
when the $\alpha^2$ effect was included in both models than when it was left out.

\subsection{Vertical advection}
The terms $-\meanv{B}\Del\cdot\meanv{U}$ (divergence) and $-\meanv{U}\cdot\Del\meanv{B}$ (advection) are approximated as
\begin{equation}
\frac{\del\mbi}{\del t}=...-\frac{\del\muz}{\del z}\mbi-\muz\frac{\del\mbi}{\del z}\simeq ...-\frac{\muz\mbi}{h}.
\end{equation}
For the flux term $-\del(\alpha\magn\muz)/\del z$ in \eqref{dalpha_mdt}, we have 
\begin{equation}
\frac{\del \alpha\magn}{\del t}=...-\frac{\alpha\magn\muz}{h}.
\end{equation}

\subsection{Approximation for $\Emf\cdot\meanv{B}$}
For $\tau\rightarrow0$, we have
\begin{equation}
\label{Emf_tau0}
\Emf\cdot\meanv{B}=c_\tau\alpha(\mean{B}_r^2+\mean{B}_\phi^2)+c_\tau\eta\turb\left(\mbr\frac{\del\mbp}{\del z}-\mbp\frac{\del\mbr}{\del z}\right),
\end{equation}
and the second bracketed term arising from the mean current helicity vanishes in the no-$z$ approximation.
Therefore, a more precise method must be used to estimate this term.
With this in mind, \citet{sss07} substituted the one-dimensional perturbation solution of the dynamo equations,
\begin{equation}
\label{Br_pert}
\mbr=R_\alpha C_0\left(\cos\frac{\pi z}{2h}+\frac{3}{4\pi}\sqrt{\frac{-D}{\pi}}\cos\frac{3\pi z}{2h}\right),
\end{equation}
\begin{equation}
\label{Bphi_pert}
\mbp=-2C_0\sqrt{\frac{-D}{\pi}}\cos\frac{\pi z}{2h},
\end{equation}
and its derivatives with respect to $z$ into Eq.~\eqref{Emf_tau0}.
The resulting expression for $\Emf\cdot\meanv{B}$ was then averaged over $0\leq z\leq h$.
In this way they obtained a non-zero estimate for $c_\tau\eta\turb(\mbr\del\mbp/\del z-\mbp\del\mbr/\del z$),
which stems from the term of order $\sqrt{-D}$ in Eq.~\eqref{Br_pert}.

In the general case of finite $\tau$, the correction must come in the evolution equations \eqref{Emfr} and \eqref{Emfphi} for $\Emfr$ and $\Emfphi$.
Averaging Eqs.~\eqref{Br_pert} and \eqref{Bphi_pert} and also their $z$-derivatives over $0\leq z\leq h$, we find
\begin{equation}
\label{dBrdz_pert}
\frac{\del\mbr}{\del z}\simeq-\frac{\pi}{2h}\left(1+\frac{3\sqrt{-D}}{4\pi^{3/2}}\right)\mbr,
\end{equation}
\begin{equation}
\label{dBpdz_pert}
\frac{\del\mbp}{\del z}=-\frac{\pi}{2h}\mbp.
\end{equation}
These expressions are used in Eqs.~\eqref{Emfr} and \eqref{Emfphi} for calculating $\Emf\cdot\meanv{B}$ in Eq.~\eqref{dalpha_mdt}.

Substituting expressions \eqref{dBrdz_pert} and \eqref{dBpdz_pert} into Eq.~\eqref{Emf_tau0}, we find for the $\tau\rightarrow0$ limit,
\begin{equation}
\label{EdotB_correc}
\Emf\cdot\meanv{B}\simeq c_\tau\alpha(\mbr^2+\mbp^2)+c_\tau\eta\turb\frac{3\sqrt{-D}}{8\pi^{1/2}h}\mbr\mbp.
\end{equation}
The term involving $\eta\turb$ used here is smaller by a factor $\pi$ than that of \citet{sss07}
since we have taken the scalar product of the $z$-averages of $\Emf$ and $\meanv{B}$, whereas those authors used the average of the scalar product.
This difference does not appear to be important.

\subsection{Testing the validity of the no-$z$ approximation}
\label{sec:noz2D}
We have performed 2D simulations in $r$-$z$ with the same parameters as our $r$-$\phi$-`no-$z$' simulation with an axisymmetric disc 
(as Models A and B but without disc flaring).
For the $r$-$z$ runs, we adopted the profiles $\alpha\kin=\mean{\alpha}\sin(\pi z/h)$, $\muz=U\f z/h$, 
as well as boundary conditions $\mbr=\mbp=0$ at $z=\pm h$.
Comparing the solution of the $r$-$z$ model (averaged over the vertical extent of the disc) with that of the $r$-$\phi$-`no-$z$' model,
we find good qualitative agreement.
We do, however, find that the saturation strength of the magnetic field is larger by a factor $\sim2$ in the $r$-$z$ model,
which suggests that this quantity is underestimated somewhat in the solutions presented in this paper.

\bibliographystyle{mn2e}
\bibliography{refs}

\label{lastpage}
\end{document}